\documentclass[aps,twocolumn,superscriptaddress, notitlepage,longbibliography, pra, 10pt]{revtex4-1}

\usepackage{amsmath} 
\usepackage{amsthm} 
\usepackage{amssymb}	
\usepackage{physics}
\usepackage[pdftex]{graphicx}
\usepackage[pdfencoding=auto, psdextra]{hyperref} 
\hypersetup{
    colorlinks,
    citecolor=blue,
    filecolor=black,
    linkcolor=red,
    urlcolor=blue
}
\usepackage{xcolor}
\usepackage{soul}
\usepackage{url}
\usepackage{makecell}
\usepackage[shortlabels]{enumitem}

\usepackage{lipsum}

\newcommand{\vbf}{\vec{\mathbf{f}}}
\newcommand{\vbu}{\vec{\mathbf{u}}}

\newcommand{\vbt}{\vec{\mathbf{t}}}

\newcommand{\vbell}{\vec{\boldsymbol{\ell}}}

\newcommand{\bH}{\mathbf{H}}

\newcommand{\bS}{\mathbf{S}}
\newcommand{\bC}{\mathbf{C}}
\newcommand{\bI}{\mathbf{I}}
\newcommand{\bK}{\mathbf{K}}
\newcommand{\bP}{\mathbf{P}}
\newcommand{\bA}{\mathbf{A}}
\newcommand{\bD}{\mathbf{D}}
\newcommand{\bT}{\mathbf{T}}

\newcommand{\Dell}{D^{(\ell)}}
\newcommand{\Du}{D^{(u)}}
\newcommand{\bDell}{\bD^{(\ell)}}
\newcommand{\bDu}{\bD^{(u)}}

\newcommand{\tp}{t^{\mathrm{ps}}}
\newcommand{\vbtp}{\vec{\mathbf{t}}^{\mathrm{ps}}}
\newcommand{\vbfLZEF}{\vbf}
\newcommand{\vbuLZM}{\vbu}
\newcommand{\vbtSSS}{\vbt}
\newcommand{\vbtZDT}{\vbt}

\newcommand{\qqc}{,\qquad}

\newtheorem{theorem}{Theorem}
\newtheorem{lemma}{Lemma}
\newtheorem{corollary}{Corollary}

\begin{document}

\title{Integrating local energetics into Maxwell-Calladine constraint counting to design mechanical metamaterials}


\author{Jason W. Rocks}
\email{jrocks@bu.edu}
\affiliation{Department of Physics, Boston University, Boston, MA 02215, USA}
\author{Pankaj Mehta}
\email{pankajm@bu.edu}
\affiliation{Department of Physics, Boston University, Boston, MA 02215, USA}
\affiliation{Biological Design Center, Boston University, Boston, MA 02215, USA}
\affiliation{Faculty of Computing and Data Science, Boston University, Boston, MA 02215, USA}

\begin{abstract}
The Maxwell-Calladine index theorem plays a central role in our current understanding of the mechanical rigidity of discrete materials. 
By considering the geometric constraints each material component imposes on a set of underlying degrees of freedom, the theorem relates the emergence of rigidity to constraint counting arguments.
However, the Maxwell-Calladine paradigm is significantly limited --
its exclusive reliance on the geometric relationships between constraints and degrees of freedom completely neglects the actual energetic costs of deforming individual components. 
To address this limitation, we derive a generalization of the Maxwell-Calladine index theorem based on susceptibilities that naturally incorporate local energetic properties such as stiffness and prestress.
Using this extended framework, we investigate how local energetics modify the classical constraint counting picture to capture the relationship between deformations and external forces.
We then combine this formalism with group representation theory to design mechanical metamaterials where differences in symmetry between local energy costs and structural geometry are exploited to control responses to external forces.
\end{abstract}

\maketitle

\section{Introduction}

Mechanical rigidity in discrete materials is an emergent property that arises from complex interactions between material components. 
In such systems, the total elastic energy is a sum of the local energy costs of deforming each discrete component.
While each local cost is a function of a component's shape,
the component shapes themselves are typically parameterized by a shared set of underlying degrees of freedom, allowing for collective interactions.
For example, in a central-force spring network, where stretching or compressing individual springs incurs an energetic cost, changes in spring length are described by the relative displacements of the network's nodes.

Within such a description, solving for the elastic response can be naturally viewed through the lens of constrained optimization -- the energetic cost of deforming each component from its minimum energy configuration imposes a soft constraint on the degrees of freedom.
Intuitively, rigidity emerges if a system is constrained enough that the degrees of freedom can no longer respond to external perturbations without an increase in energy. 
However, formalizing this intuition into a general theory of rigidity for discrete systems remains an open problem.

As implementations of this idea, Maxwell's rule for rigidity~\cite{Maxwell1864} and its successor, the Maxwell-Calladine index theorem~\cite{Calladine1978, Lubensky2015}, represent important milestones in our understanding of rigidity. 
In any system well-approximated by a spring network,
the Maxwell-Calladine theorem reduces the assessment of linear stability to an exercise in counting constraints and degrees of freedom.
Originally developed in the context of mechanical frame assemblies, this constraint counting argument is central to theories of mechanical stability for a variety of systems including structural glasses~\cite{Phillips1981, Boolchand2005}, jammed packings of spheres~\cite{Wyart2005, vanHecke2010, Liu2010}, and biopolymer networks~\cite{Broedersz2014}, and has even been extended to systems with more complicated interactions such as bond-bending dominated networks~\cite{Rens2019}.
Furthermore, this index theorem has played a crucial role in the development of mechanical metamaterials with topologically protected boundary modes, analogous to those observed in electronic quantum systems~\cite{Kane2014, Bertoldi2017}.

Despite the considerable insights they yield, constraint counting arguments based on the Maxwell-Calladine paradigm are fundamentally incomplete.
Because they rely exclusively on the linear geometric relationships between the component shapes and degrees of freedom, they fail to capture rigidity in instances where local energetic properties -- such as stiffness and prestress -- or nonlinearities in the geometry play an important role.
In particular, they cannot account for the stabilizing effects of prestress, which interacts with nonlinear aspects of the geometry to affect the material response even in the linear regime~\cite{Calladine1991, Connelly1996, Connelly2022}.
Because this phenomenon is a basic feature of materials throughout structural engineering, physics, and biology~\cite{Zhang2021}, a general theory of rigidity that incorporates prestress could prove invaluable.
To this end, various alternative rigidity criteria have been proposed~\cite{Connelly1996, Zhang2021, Damavandi2022, Damavandi2022a}. 
However, these criteria do not explicitly take into consideration the precise details of the energetic properties, but instead take an indirect approach,
focusing on the second-order features of the geometry that can interact with prestress under the right circumstances.

Here, we propose a description of mechanical rigidity in general discrete systems based on the susceptibility to external perturbations.
By considering the elastic responses of both the discrete components and their underlying degrees of freedom to their respective conjugate external forces, 
we derive a generalization of the Maxwell-Calladine index theorem that naturally incorporates energetic properties. 
Under this framework, we show that the collective modes of the original Maxwell-Calladine picture -- linear zero modes (LZMs) and states of self-stress (SSSs) -- are each related to specific types of external perturbations, which we call linear zero-extension forces (LZEFs) and zero-displacement tensions (ZDTs).
Combining this formalism with classic results from group representation theory, we explore how different aspects of a material's symmetry can dictate the relationship between these different types of modes.
Using simple examples, we then demonstrate how this relationship may be exploited to design metamaterials with specific responses to external perturbations, focusing on materials where local energy costs break or preserve symmetries in the geometry.

\subsection{Organization of paper}

In Sec.~\ref{sec:theory}, we start by introducing the theoretical background we will use to explore elasticity in discrete materials and provide a brief derivation of the classic Maxwell-Calladine index theorem.
In Sec.~\ref{sec:suscept}, we introduce a set of elastic susceptibilities describing responses to different types of external perturbations and explore the taxonomy of their collective mode structures.
In Sec.~\ref{sec:GIT}, we use these susceptibilities to derive a generalization of Maxwell-Calladine that includes energetic properties like stiffness and prestress.
In Sec.~\ref{sec:collective}, we explore the relationships between the modes of the classic Maxwell-Calladine index theorem and our generalized version and formalize this relationship as a mathematical theorem.
In Sec.~\ref{sec:sym}, we use group representation theory to prove a second theorem describing how symmetry affects the relationships between these modes.
We then apply these results to provide simple examples where prestress is used to control responses to external forces by manipulating symmetry.
Finally, in Sec.~\ref{sec:discussion}, we conclude and discuss the implications of our results for understanding and designing mechanical metamaterials.

\begin{figure*}[t]
\centering
\includegraphics[width=\linewidth]{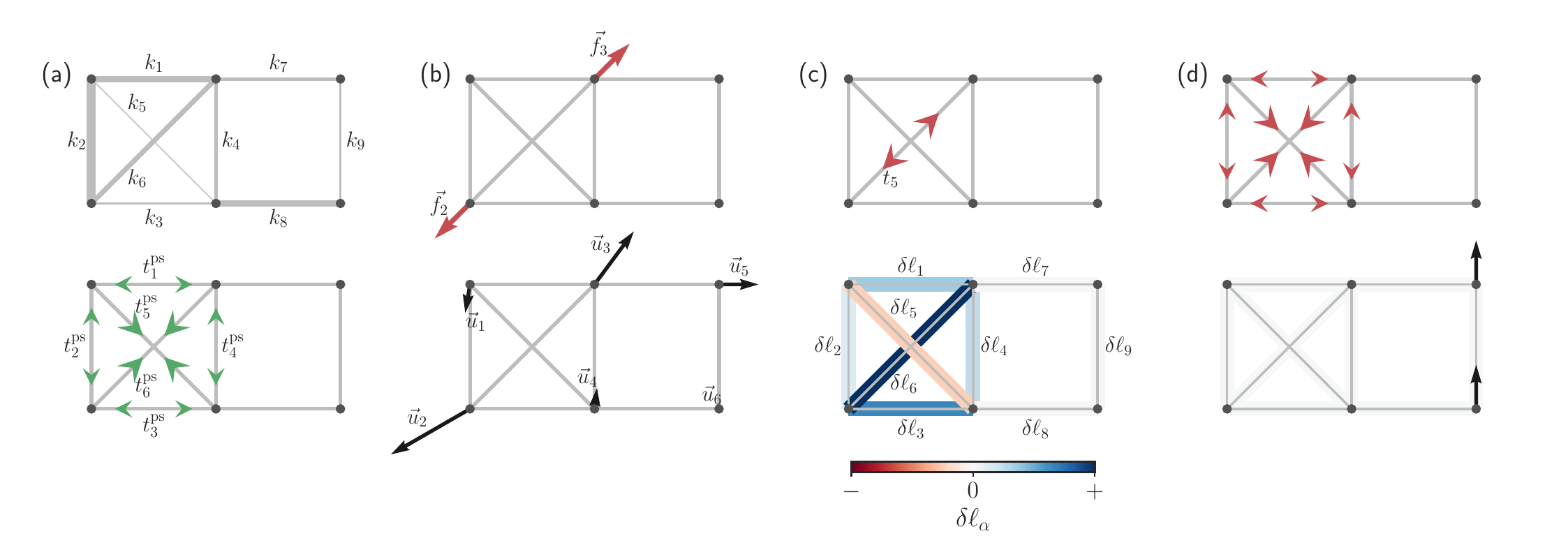}
\caption{\textbf{Local energetic properties, external perturbations, and elastic responses.}
The linear elastic properties are shown for a simple central-force spring network.
\textbf{(a)} The stiffnesses of the bonds $k_\alpha$ (top), represented by line thickness, characterize their compressibility, while the prestresses $\tp_\alpha$ (bottom), characterize any geometric frustration that arises when bonds are stretched or compressed in the ground state configuration.
Green arrows indicate the prestress tension exerted by each bond on the rest of the network.
 \textbf{(b)} From the perspective of the nodes, external forces $\vec{f}_i$ (top) result in displacements $\vec{u}_i$ (bottom).
The notation $\vec{u}_i$ and $\vec{f}_i$ represent the $d$-dimensional vectors of displacements and forces associated with each node $i$ for a network in $d$ dimensions.
 \textbf{(c)} From the perspective of the bonds, the external forces on the nodes are equivalent to external bond tensions $t_\alpha$ (top), while node displacements correspond to bond extensions $\delta \ell_\alpha$ (bottom) indicated by the color scale.
\textbf{(d)} The network contains a state of self-stress (top) [identical to the prestress in (a)] composed of external tensions that result in no net forces on the nodes, along with a linear zero mode (bottom) composed of displacements do result in zero extension of the bonds.
 }
\label{fig:defs}
\end{figure*}

\section{Elasticity Theory for Discrete Materials}\label{sec:theory}

In this section, we introduce the basic concepts and formalism we will use in the remainder of the paper. 
We consider a mechanical system consisting of $N_\mathrm{c}$ discrete components, each with size $\ell_\alpha(\{u_i\})$ (${\alpha=1,\ldots,N_\mathrm{c}}$),
 parameterized in terms of $N_{\mathrm{dof}}$ degrees of freedom $u_i$ (${i=1,\ldots,N_{\mathrm{dof}}}$). For convenience, we organize these quantities into the vectors $\vbell$ and $\vbu$, respectively.
To evoke a concrete image, we will often discuss these quantities in the context of central-force spring networks in which the degrees of freedom are associated with the nodes, while the components are the springs, or bonds, defined by the edges of the network.
For such a spring network with $n$ nodes in $d$ dimensions, $\vbu$ is a $dn$-dimensional vector of node displacements, while $\vbell$ is a $N_\mathrm{c}$-dimensional vector of bond lengths.
Accordingly, we often refer to $\vbu$ and $\vbell$ as generalized ``node displacements'' and ``bond lengths,'' respectively.

Beyond central-force spring networks~\cite{Lubensky2015}, a broad range of systems in soft matter and biological physics may be cast in the above form.
For example, in vertex models of biological tissues the generalized lengths $\vbell$ may refer to the areas and perimeters of cells in a two-dimensional epithelial tissue, or the cell volumes, surface areas, and edge lengths in a more general three-dimensional tissue~\cite{Alt2017}.
Alternatively, the discrete components do not need to correspond to physical objects, but instead may be abstract in nature.
Such examples include the lengths of bonds between neighboring particles in a jammed packing~\cite{Liu2010}, the angles associated with bond-bending costs between adjacent fiber segments in a  biopolymer network~\cite{Broedersz2014},
or the angles between connected facets in an origami structure~\cite{Schenk2016}.
Similarly, the displacements $\vbu$ may describe the change in any type degree of freedom from a system's ground state, such as the orientational angles of particles that lack spherical symmetry like in jammed packings of elliptical particles~\cite{Donev2007, Mailman2009}.

\subsection{Energy and external perturbations}

We now write down the generalized Hamiltonian for these discrete materials. Defining $\vbu = 0$ as the system's ground state in the absence of external perturbations, we measure changes in length via a set of generalized ``bond extensions,'' $\delta \vbell(\vbu) \equiv \vbell(\vbu) - \vbell_0$, where $\vbell_0 \equiv \vbell(0)$ are the ``equilibrium lengths'' of the bonds in the ground state. 

At zero temperature, we consider the Hamiltonian
\begin{equation}
\mathcal{H}(\vbu) = E(\vbell(\vbu)) + \frac{\lambda}{2}\norm*{\vbu}^2 - \vbf\cdot \vbu - \vbt \cdot \delta \vbell(\vbu),\label{eq:hamiltonian}
\end{equation}
where the elastic energy $E(\vbell(\vbu))$ is expressed as a sum of the energetic costs of deforming each bond $\alpha$,
\begin{equation}
E(\vbell(\vbu)) = \sum_\alpha V_\alpha(\ell_\alpha(\vbu);\ell^*_\alpha).\label{eq:energy}
\end{equation}
We introduce two externally applied generalized forces, $\vbf$ and $\vbt$, conjugate to the displacements $\vbu$ and bond lengths $\delta \vbell(\vbu)$, respectively.
By convention, we will often refer to $\vbf$ as the external force and $\vbt$ as the external tension.
In general, we will be interested in linear response theory where these external perturbations are small. 

We also add a small harmonic regularization term proportional to $\lambda$ which penalizes large displacements. This regularization serves as a convenient bookkeeping tool to keep track of deformations with no energetic cost (i.e., mechanisms or zero-energy modes). We will eventually be interested in the limit $\lambda\rightarrow 0$ where this term becomes negligible.

By convention, we define the rest length $\ell^*_\alpha$ for each bond such that ${V_\alpha(\ell^*_\alpha; \ell^*_\alpha) = 0}$ (in general, $\ell^*_\alpha$ may differ from $\ell_{0,\alpha}$).
For example, in a harmonic spring network, the potentials take the form ${V_\alpha(\ell_\alpha(\vbu); \ell^*_\alpha) = \frac{1}{2}k_\alpha(\ell_\alpha(\vbu) - \ell^*_\alpha)^2}$ where $k_\alpha$ is the stiffness of bond $\alpha$.

To determine the elastic response, we minimize Eq.~\eqref{eq:hamiltonian} with respect to the displacements by setting its gradient to zero,
\begin{equation}
0 = \pdv{\mathcal{H}(\vbu)}{u_i},\label{eq:gradient}
\end{equation}
and then solve the resulting equations for the displacements $\vbu$.

\subsection{Local energetic properties in the linear regime}

In the linear regime, the elastic energy [Eq.~\eqref{eq:energy}] close to the ground state is characterized by the Hessian matrix
\begin{equation}
\begin{aligned}
H_{ij} &= \pdv{E}{u_i}{u_j}\\
& =  \sum_{\alpha\beta}\pdv{\ell_\alpha}{u_i}\pdv{E}{\ell_\alpha}{\ell_\beta}\pdv{\ell_\beta}{u_j} +  \sum_\alpha\pdv{E}{\ell_\alpha} \pdv{\ell_\alpha}{u_i}{u_j}.
\end{aligned}
\end{equation}
In matrix form, we will use the notation
\begin{equation}
\bH = \bC \bK \bC^T + \bT\label{eq:Hessian}
\end{equation}
where we introduce the compatibility matrix $\bC$ (or alternatively, the equilibrium matrix $\bC^T$) and prestress matrix $\bT$ with elements
\begin{gather}
C_{\alpha i} = \pdv{\ell_\alpha}{u_i}\label{eq:compat}\\
T_{ij} = \sum_\alpha \tp_\alpha \pdv{\ell_\alpha}{u_i}{u_j}.
\end{gather}
We also define the prestress vector $\vbtp$ and stiffness matrix $\bK$, 
\begin{gather}
\tp_\alpha = \pdv{E}{\ell_\alpha} = V'_\alpha,\label{eq:prestress}\\
K_{\alpha\beta} = \pdv{E}{\ell_\alpha}{\ell_\beta} = V''_\alpha \delta_{\alpha\beta} = k_\alpha\delta_{\alpha\beta}.\label{eq:stiffness}
\end{gather}
These two quantities encode the system's local energetic properties in the linear regime. While the stiffness $k_\alpha$ describes the energetic cost of deforming bond $\alpha$, the prestress $\tp_\alpha$ describes any frustration arising due to incompatibility between a bond's rest length and its equilibrium length in the system's global ground state (${\ell^*_\alpha\neq \ell_{0,\alpha}}$). 
As an example, in a harmonic spring network, prestress takes the form ${\tp_\alpha = k_\alpha(\ell_{0,\alpha} - \ell^*_\alpha)}$. We note that all the above quantities are evaluated at the system's ground state, $\vbu = 0$.

In Fig.~\ref{fig:defs}(a), we demonstrate possible values of the stiffness and prestress for a simple central-force harmonic spring network.
To achieve the prestress depicted in this ground state configuration, we assign incompatible rest lengths to bonds 1-6  so that the four outer bonds (1-4) are compressed in the ground state, $\ell_{0,\alpha} < \ell_\alpha^*$, while the two diagonal bonds (5-6) are stretched, $\ell_{0,\alpha} > \ell_\alpha^*$.
This creates frustration that cannot be relaxed away by moving the nodes.

\subsection{Node space and bond space perspectives}

In general, it is helpful to think about the linear response in two different spaces, node space (degrees of freedom) and bond space (components or constraints). 
In node space, applying a set of forces $\vbf$ to the nodes results in a set of displacements $\vbu$. 
An example of this is shown in Fig.~\ref{fig:defs}(b), where external applied forces (top) result in displacements of the nodes (bottom).
Alternatively, we can think about how these quantities in node space correspond to their counterparts in bond space, tensions, and extensions.
Fig.~\ref{fig:defs}(c) depicts a set of bond tensions $\vbt$ (top) that result in the same set of displacements as the forces in Fig.~\ref{fig:defs}(b), along with the bond extensions $\delta \vbell$ (bottom) resulting from those displacements.
Intuitively, it is clear that every set of external tensions is equivalent to some set of external forces, while each set of displacements results in a set of bond extensions.

In the linear regime, the correspondence between quantities in node space and bond space is provided by the compatibility matrix $\bC$ 
via the relations
\begin{equation}
\delta \vbell = \bC\vbu\qqc\vbf = \bC^T \vbt \label{eq:Crelations}.
\end{equation} 
The first relation between the extensions and displacements is simply a result of the chain rule. The second relation between external tensions and forces requires a slightly more subtle argument.
Expanding the tension term in Eq.~\eqref{eq:hamiltonian} to linear order in $\vbu$, we get
\begin{equation}
\vbt \cdot \delta \vbell(\vbu) \approx \sum_i \vbt \cdot \pdv{\vbell}{u_i}\eval_{\vbu=0} u_i =(\bC^T \vbt ) \cdot \vbu.
\end{equation}
Comparing this expression with the force term in Eq.~\eqref{eq:hamiltonian}, $\vbf\cdot \vbu$,
we see that applying an external tension $\vbt$ to the bonds is equivalent to applying a force ${\vbf =\bC^T \vbt}$ to the nodes.  

\subsection{The simplest case: zero prestress with uniform stiffness}\label{sec:simple}

Throughout this paper, it will be instructive to examine the elastic response in the particularly simple case of zero prestress $\vbtp=0$ with uniform unit bond stiffness ${\bK = \bI}$. 
It is within this setting that the Maxwell-Calladine framework is typically derived~\cite{Lubensky2015}.
In this special case, the Hessian [Eq.~\eqref{eq:Hessian}] attains the simple form
\begin{equation}
\bH = \bC \bC^T\label{eq:simpleHessian}.
\end{equation}
Essentially, when the energetic properties are trivial, the Hessian contains the same information as the compatibility matrix $\bC$.
In other words, deformations are completely characterized by the system's geometry encoded in $\bC$.
However, in more general situations -- such as in the presence of prestress or heterogeneous bond stiffnesses -- the Hessian and its associated elastic response will also depend on the precise details of the energetic properties encoded in  $\bK$ and $\vbtp$.

\subsection{Maxwell-Calladine Index Theorem}\label{sec:MCIT}

We now present a brief derivation of the classic Maxwell-Calladine index theorem with an eye to generalizing it to include prestress and heterogeneous bond stiffness.
In the Maxwell-Calladine framework, a central role is played by the compatibility matrix $\bC$. 
The compatibility matrix relates quantities that reside in the $N_{\mathrm{c}}$-dimensional bond space to those that live in the $N_{\mathrm{dof}}$-dimensional node space.
According to Eq.~\eqref{eq:Crelations},  $\bC$ is formally a linear map from displacements in node space $\vbu$ to extensions in bond space $\vbell$, while $\bC^T$ is a linear map from tensions $\vbt$ in bond space to forces in node space $\vbf$.

Applying rank-nullity theorem, we can express the dimension of the domain of each operator as the sum of its rank plus the dimension of its kernel (right null space).
According to Eq.~\eqref{eq:Crelations}, the kernel of $\bC$ is composed of linear zero modes (LZMs), or displacements that do not extend or compress the bonds to linear order. Thus, we have
\begin{equation}
N_{\mathrm{dof}} = \mathrm{rank}(\bC) + \mathrm{\# LZMs}.
\end{equation}
Analogously, the kernel of $\bC^T$ is composed of states of self-stress (SSSs), or tensions that balance to create zero net forces on the nodes, and we have
\begin{equation}
N_{\mathrm{c}} = \mathrm{rank}(\bC^T) + \mathrm{\# SSSs}.
\end{equation}
Because $\bC$ and $\bC^T$ are transposes of one another, their ranks are equal. 
Using this observation, we subtract the second equation from the first, yielding the Maxwell-Calladine index theorem,
\begin{equation}
 N_{\mathrm{dof}} - N_{\mathrm{c}}  = \mathrm{\# LZMs} - \mathrm{\# SSSs},\label{eq:MCIT}
\end{equation}
which relates the difference in the number of degrees of freedom and constraints to the difference in the number of LZMs and SSSs.
Based on Eq.~\eqref{eq:MCIT}, one concludes a system is rigid to first order when the number of LZMs is zero (excluding global translations and rotations, if applicable).

As an illustration, we analyze the rigidity of the network in Fig.~\ref{fig:defs}, 
which has six nodes in two dimensions for a total of $N_{\mathrm{dof}}=12$ degrees of freedom and $N_{\mathrm{c}} = 9$ bonds that impose constraints on the system.
This network also has one SSS, shown in the top of Fig.~\ref{fig:defs}(d), which happens to be identical to the prestress in Fig.~\ref{fig:defs}(b) (the prestress $\vbtp$ is always a SSS since in the ground state ${0 = \partial E/\partial \vbu = \bC^T\vbtp}$).
Solving Eq.~\eqref{eq:MCIT}, we find that this network should have four LZMs.
Since two of these modes correspond to global translations along the $x-$ and $y-$axes and one to a global rotation, there must be one additional LZM, shown in the bottom of Fig.~\ref{fig:defs}(d).
Thus, we conclude that the system is not rigid to first order.

\section{Elastic Susceptibilities}\label{sec:suscept}

The derivation of Maxwell-Calladine is based exclusively on the geometry encoded in the compatibility matrix $\bC$, ignoring prestress and stiffness.  
To incorporate these energetic properties and derive a generalized index theorem, it is necessary to assess a system's rigidity directly via its response to external perturbations, rather than indirectly through the geometry. 
The response of a discrete material can be characterized by four distinct susceptibility matrices that measure how displacements and extensions respond to externally applied forces and tensions. These susceptibilities play a central role in what follows and are the key physical objects necessary to generalize the Maxwell-Calladine framework to include energetics. 

\subsection{Definitions and expressions for susceptibilities}

The natural objects of all linear response theories are susceptibilities that characterize how the system responds to external perturbations.
As hinted by our choice of Hamiltonian in Eq.~\eqref{eq:hamiltonian}, we will focus on the response of the system to the application of external forces $\vbf$ and tensions $\vbt$. 
The response can be characterized either in terms of the displacements $\vbu$ or extensions $\delta \vbell$.
This results in four distinct yet related matrix susceptibilities:
\begin{equation}
\begin{aligned}
\qty[\pdv{\vbu}{\vbf}]_{ij} &= \pdv{u_i}{f_j} ,& \qty[\pdv{\vbu}{\vbt}]_{i\beta} &= \pdv{u_i}{t_\beta},\\
\qty[\pdv{\vbell}{\vbf}]_{\alpha j} &= \pdv{\ell_\alpha}{f_j}, & \qty[\pdv{\vbell}{\vbt}]_{\alpha\beta} &= \pdv{\ell_\alpha}{t_\beta}.
\end{aligned}
\end{equation}
It will be helpful to subdivide these four susceptibilities into the two square ``diagonal'' susceptibilities, ${\partial \vbu / \partial\vbf}$ and ${\partial \vbell / \partial\vbt}$ , which map between spaces of the same dimensions, and the two rectangular ``off-diagonal'' susceptibilities, ${\partial \vbu / \partial\vbt}$ and ${\partial \vbell / \partial\vbf}$, that map between spaces of different dimensions.

To derive explicit expressions for these susceptibilities, we assume that we are looking at perturbations of stable minimum energy configurations so that the Hessian $\bH$ is positive semi-definite.
We differentiate the $N_{\mathrm{dof}}$ equations of Eq.~\eqref{eq:gradient} with respect to $f_j$ and solve the resulting matrix equations to obtain an explicit expression for ${\partial \vbu / \partial\vbf}$. 
The other susceptibilities are then derived from ${\partial \vbu / \partial\vbf}$ using the chain rule. 
As shown in Appendix~\ref{app:derivation}, in the limit $\lambda\rightarrow 0$, 
the four susceptibilities are
\begin{equation}
\begin{aligned}
\pdv{\vbu}{\vbf} &= \bH^+ + \frac{1}{\lambda}\bP_H^{\mathrm{ZEM}}, & \pdv{\vbu}{\vbt} &= \pdv{\vbu}{\vbf}\bC^T,\\
\pdv{\vbell}{\vbf} &= \bC\pdv{\vbu}{\vbf}, & \pdv{\vbell}{\vbt} &= \bC\pdv{\vbu}{\vbf}\bC^T,\label{eq:suscept}
\end{aligned}
\end{equation}
where $\bH^+$ is the Moore-Penrose inverse, or pseudoinverse, of the Hessian and ${\bP^{\mathrm{ZEM}}_H = \bI - \bH\bH^+}$ is an orthogonal projector onto its kernel, containing the zero-energy modes (ZEMs) of the Hessian. As a reminder, the pseudoinverse provides a generalization of matrix inverses to singular or rectangular matrices by only inverting the nonsingular part of a matrix (see Sec.~\ref{app:pseudoinverse} of the Appendix for a detailed definition).

\begin{figure}[t!]
\centering
\includegraphics[width=\linewidth]{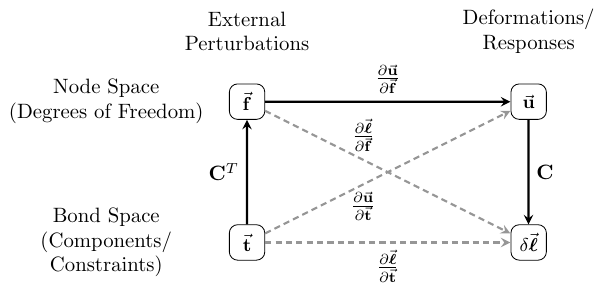}
\caption{\textbf{Susceptibilities and their relationships as maps between vector spaces.} The four susceptibilities, the compatibility matrix, and its transpose provide maps between forces, tensions, displacements, and extensions. While, $\bC$ and $\bC^T$ map between node space (degrees of freedom) (top row) and bond space (components/constraints) (bottom row), the susceptibilities map external perturbations (left column) to deformation responses (right column).
Arrows indicate the direction of each map with dashed arrows representing operators that are compositions of other operators.
}
\label{fig:maps}
\end{figure}

\subsection{Susceptibilities as maps between vector spaces}

\begin{table*}[t!]
\centering
\caption{
Collective Mode Taxonomy
}
\label{tab:tax}
\begin{tabular}{c@{$\quad:\quad$}c@{$\quad\rightarrow\quad$}c|c|c}
\hline\hline
Operator & Domain & Codomain & Kernel Modes & Abbreviation\\
\hline
$\bC$ & Displacements $\vbu$ & Extensions $\delta \vbell$ &  Linear Zero Modes & LZM\\
$\bC^T$ & Tensions $\vbt$ & Forces $\vbf$ & States of Self-Stress & SSS\\\hline
$\pdv{\vbu}{\vbt}$ &  Tensions $\vbt$ & Displacements $\vbu$ &  Zero-Displacement Tensions & ZDT\\
$\pdv{\vbell}{\vbf}$ &  Forces $\vbf$ & Extensions $\delta \vbell$ & Linear Zero-Extension Forces & LZEF\\
\hline
$\bH$ & Displacements $\vbu$ & Forces $\vbf$ & Zero-Energy Modes & ZEM\\
\hline\hline
\end{tabular}
\end{table*}

In our formalism, a privileged role is played by the diagonal susceptibility ${\partial \vbu / \partial\vbf}$. Physically, ${\partial \vbu / \partial\vbf}$ is linear map from external forces to the resulting displacements of the nodes. To gain intuition about this susceptibility, it is helpful to examine the explicit expression for ${\partial \vbu / \partial\vbf}$ in Eq.~\eqref{eq:suscept}. The expression is composed of two parts: the pseudoinverse of the Hessian $\bH^+$ and the projection operator $\bP_H^{\mathrm{ZEM}}$. The pseudoinverse $\bH^+$ simply describes harmonic motion of the degrees of freedom around their minimum energy positions in rigid directions (i.e., directions in which the Hessian has nonzero eigenvalues). On the other hand, the projection operator $\bP_H^{\mathrm{ZEM}}$ describes motions resulting from exciting modes that cost zero energy (i.e., eigenmodes of the Hessian with eigenvalues of zero). Furthermore, notice that the projector term is proportional to $\lambda^{-1}$. Thus, in the physical limit without regularization, $\lambda \rightarrow 0$, forces that excite these zero-energy modes result in unconstrained displacements with magnitudes that diverge as $\lambda^{-1}$.

More generally, the four susceptibilities in Eq.~\eqref{eq:suscept} combine with $\bC$ and $\bC^T$ to form a set of linear operators that can be composed to map either type of external perturbation ($\vbf$ or $\vbt$) to either type of deformation ($\vbu$ or $\delta \vbell$). We illustrate their relationships graphically in Fig.~\ref{fig:maps}. According to Eq.~\eqref{eq:Crelations}, $\bC$ relates deformations in node space (displacements) to those in bond space (extensions), and $\bC^T$ relates external perturbations in bond space (tensions) to those in node space (forces).
In contrast, the susceptibilities always relate external perturbations to deformations.  
In this picture, $\bC$, $\bC^T$, and ${\partial \vbu / \partial\vbf}$ can be viewed as a set of ``fundamental'' operators (shown with solid arrows) whose compositions give rise to the remaining three susceptibilities (shown with dashed arrows).

\subsection{Collective mode taxonomy}

Just as the original Maxwell-Calladine theorem relates collective modes in the kernels of $\bC$ and $\bC^T$, our generalized index theorem will relate modes in the kernels of ${\partial \vbu / \partial\vbt}$ and ${\partial\vbell / \partial\vbf}$. For this reason, it is worth briefly summarizing our nomenclature for these collective modes, as well as their physical meanings. As discussed earlier in the context of the Maxwell-Calladine theorem, the spaces spanned by the kernels of $\bC$ and $\bC^T$ are usually referred to as linear zero modes (LZMs) and states of self-stress (SSS), respectively. The Maxwell-Calladine theorem relates the dimensions of these two spaces to the number of degrees of freedom and constraints.

In our generalized mode taxonomy, an analogous role will be played by the modes that span the kernels of ${\partial \vbu / \partial\vbt}$ and ${\partial\vbell / \partial\vbf}$. We note that both of these types of modes correspond to \emph{external perturbations}. 
Physically, modes that lie in the kernel of ${\partial\vbell / \partial\vbf}$ correspond to nontrivial sets of external forces $\vbf$ that may displace the nodes, but do not actually extend the bonds. 
We will refer to these modes as \emph{linear zero-extension forces} (LZEFs). 
Similarly, modes that lie in the kernel of ${\partial \vbu / \partial\vbt}$ correspond to sets of external tensions that do not result in any displacement of the nodes when applied to the bonds. 
We will refer to these modes as \emph{zero-displacement tensions} (ZDTs).
See Table~\ref{tab:tax} for a summary of this collective mode taxonomy.

\section{Generalized Index Theorem}\label{sec:GIT}

We are now in a position to derive a generalization of the Maxwell-Calladine index theorem. Importantly, our generalized index theorem holds even in the presence of nonzero prestress or heterogeneous bond stiffnesses. 
We focus on the two off-diagonal susceptibilities relating node and bond space, ${\partial \vbu / \partial\vbt}$ and ${\partial \vbell / \partial\vbf}$.
Following a procedure analogous to the derivation in Sec.~\ref{sec:MCIT}, 
we apply rank-nullity theorem to express the dimension of the domain of each operator as a sum of its rank plus the dimension of its kernel.
Using the definitions of LZEFs and ZDTs from the previous section as the collective modes in the kernels of the two susceptibilities, we obtain
\begin{align}
N_{\mathrm{dof}} &= \mathrm{rank}\qty(\pdv{\vbell}{\vbf}) + \mathrm{\# LZEFs},\\
N_{\mathrm{c}} &= \mathrm{rank}\qty(\pdv{\vbu}{\vbt}) + \mathrm{\# ZDTs}.
\end{align}
Since the two susceptibilities are transposes of one another [see Eq.~\eqref{eq:suscept}], their ranks are equal. Using this fact, we subtract the second equation from the first to arrive at our generalized index theorem,
\begin{equation}
 N_{\mathrm{dof}} - N_{\mathrm{c}} = \mathrm{\# LZEFs} - \mathrm{\# ZDTs}.\label{eq:genMC}
\end{equation}
Comparing to the classic version in Eq.~\eqref{eq:MCIT},
we see that the left-hand side is identical, containing the difference in the number of degrees of freedoms and constraints,
while the right-hand side now contains the difference in the number of LZEFs and ZDTs.
At first glance, these modes seem to be very different from those usually considered in the Maxwell-Calladine theorem, namely LZMs and SSSs. 
Whereas LZMs and SSSs are purely properties of the system's geometry encoded by $\bC$ and $\bC^T$,
LZEFS and ZDTs are types of external perturbations that depend on both the system's geometry and local energetics (prestress and stiffness).
We will see that these two sets of modes are intimately related and can even be shown to be equivalent in special cases depending on the precise details of the energy.

\section{Collective Mode Relationships}\label{sec:collective}

In this section, we explore the relationship between the collective modes of the Maxwell-Calladine index theorem (SSSs and LZMs) and their counterparts in the generalized version (ZDTs and LZEFs).
We first demonstrate this relationship explicitly in the  simple case discussed in Sec.~\ref{sec:simple} where the details of the energetic properties can essentially be ignored (zero prestress and uniform bond stiffness).
We then move on to the general case of arbitrary local energetic properties where we formalize these results into a theorem.

\subsection{The simplest case: zero prestress with uniform stiffness}

In the case of zero prestress, $\vbtp = 0$, with uniform unit stiffness, $\bK = \bI$, 
the susceptibility ${\partial \vbu / \partial\vbf}$ takes on an especially simple form.
Combining Eqs.~\eqref{eq:simpleHessian} and ~\eqref{eq:suscept}, and using properties of the pseudoinverse, we obtain
\begin{equation}
\pdv{\vbu}{\vbf} = \qty[\bC^T\bC]^+ + \frac{1}{\lambda}\qty[\bI - \bC^+ \bC],\label{eq:simpledufdf}
\end{equation}
where to reiterate, the first term is the pseudoinverse of the Hessian and the second term is a projector onto the zero-energy modes.
According to Eq.~\eqref{eq:suscept}, we then appropriately multiply by $\bC$ or $\bC^T$  to obtain simplified forms for the two off-diagonal susceptibilities,
\begin{equation}
\pdv{\vbu}{\vbt} = \bC^+\qqc  \pdv{\vbell}{\vbf} = \qty[\bC^+]^T.\label{eq:simplesuscepts}
\end{equation}
When taking the pseudoinverse of a general rectangular matrix such as $\bC$, the row and column spaces of the matrix are swapped, along with the left and right null spaces (see Appendix~\ref{app:pseudoinverse}). 
Thus, the kernel of ${\partial \vbu / \partial\vbt}$ is the same as $\bC^T$ and similarly, the kernel of ${\partial\vbell / \partial\vbf}$ is the same as $\bC$.
In other words, for the simple case of zero prestress with uniform stiffness, the two index theorems provide identical information: the SSSs and LZMs are the same as the ZDTs and LZEFs, respectively!

To make this connection more transparent, we take advantage of the fact that the susceptibility ${\partial \vbu / \partial\vbf}$ is guaranteed to be full rank and is therefore invertible.
While ${\partial \vbu / \partial\vbf}$ uniquely maps external forces to their resulting displacements, its inverse $[\partial \vbu / \partial\vbf]^{-1}$ uniquely maps displacements to the external forces that created them.

Now consider a LZEF $\vbfLZEF$. Since LZEFs do not extend the bonds, the displacement created by $\vbfLZEF$ must preserve the bond lengths as well.
Thus, every $\vbfLZEF$ results in a displacement  $\vbuLZM$ along a LZM given by
\begin{equation}
\vbuLZM = \pdv{\vbu}{\vbf}\vbfLZEF.\label{eq:LZMLZEF}
\end{equation}
Because ${\partial \vbu / \partial\vbf}$ is invertible, we may also solve the above equation for a fixed $\vbuLZM$ to find the unique $\vbfLZEF$ that couples to it.
In short, the LZMs and LZEFs are isomorphic, related by the map provided by ${\partial \vbu / \partial\vbf}$.

For the case of zero prestress and uniform stiffness, Eq.~\eqref{eq:LZMLZEF} is especially simple.
Using Eqs.~\eqref{eq:simpledufdf} and \eqref{eq:simplesuscepts}, combined with the fact that $\vbf$ lies in the kernel of ${\partial \vbell / \partial \vbf}$, we find
\begin{equation}
\vbuLZM = \frac{1}{\lambda}\vbfLZEF.\label{eq:LZMLZEFzimple}
\end{equation}
Thus, in this simple case, each LZEF couples to a LZM that is the same up to a constant prefactor.
Furthermore, the prefactor of $\lambda^{-1}$ indicates that in this case, all LZMs also happen to be zero-energy modes.

The relationship between ZDTs and SSSs is even more straightforward since they are both types of external tensions.
Now note that by definition, a SSS does not generate net forces on the nodes.
Because ${\partial \vbu / \partial\vbf}$ is full rank and uniquely maps from forces to displacements,
the zero net force created by a SSS must also result in zero displacement of the nodes.
Similarly, because a ZDT does not generate displacements, applying $[\partial \vbu / \partial\vbf]^{-1}$ maps the zero displacement created by a ZDT to zero net forces on the nodes.
Thus, we conclude that every SSS is a ZDT and every ZDT is a SSS. 
Evidently, the two types of modes are always the same.

\subsection{Collective Mode Correspondence Theorem}
We formulate the observations from the previous section into the following theorem (proof provided in Appendix~\ref{app:proofs}):
\begin{theorem}\label{thm:isomorphic}
Consider a discrete elastic system described by the Hamiltonian in Eq.~\eqref{eq:hamiltonian}. The following statements are true:
\begin{enumerate}[(i)]
\item The ZDTs and SSSs are equivalent.
\item The LZEFs are isomorphic to the LZMs, with each LZEF $\vbfLZEF$ coupling to a unique LZM $\vbuLZM$ according to the bijective map
\begin{align}
\vbuLZM = \pdv{\vbu}{\vbf}\vbfLZEF.\label{eq:isomorphism}
\end{align}
\end{enumerate}
\end{theorem}
We emphasize that in general,
the LZEFs and LZMs span different subspaces of the $N_{\mathrm{dof}}$-dimensional node space.
The precise correspondence between these two types of collective modes depends on the interplay of the system's geometric structure and its local energetic properties encoded in the susceptibility ${\partial \vbu / \partial\vbf}$ (which is just the Hessian in disguise).

In the previous section, we saw that for zero prestress and uniform stiffness, the LZEFs and LZMs were identical -- each LZEF coupled to a displacement along a parallel LZM [Eq.~\eqref{eq:LZMLZEFzimple}].
While this case is especially simple, this mode equivalence actually generalizes to a much broader class of systems, described by the following corollary to Theorem.~\ref{thm:isomorphic}:
\begin{corollary}\label{cor:eigenmode}
Consider a LZM $\vbuLZM$ and a LZEF $\vbfLZEF$ related by Eq.~\eqref{eq:isomorphism}. The two modes are parallel, $\vbuLZM \parallel \vbfLZEF$, if and only if they are eigenmodes of Hessian.
\end{corollary}
We explore some implications of this statement for the design of metamaterials in the next section.

\section{Symmetry Breaking with Local Energetic Properties}\label{sec:sym}

In this section, we explore the role of symmetry in controlling the relationship between LZMs and LZEFs described in the previous section.
Using group representation theory, we show how a system's geometric and energetic symmetries combine to control the isomorphism in Theorem~\ref{thm:isomorphic} between the two types of modes, and ultimately determine whether they are eigenmodes of the Hessian as described in Corollary~\ref{cor:eigenmode}.
Using simple examples, we then demonstrate how this idea may be exploited to design metamaterials with specific responses.

\subsection{Collective Mode Symmetry Theorem}

Recall that LZMs are defined as modes in the kernel of the compatibility matrix $\bC$, 
while LZEFs are defined as modes in the kernel of the susceptibility ${\partial\vbell / \partial\vbf}$.
For these two types of modes to match, their operators must share at least part of their mode structures (i.e., be at least partially simultaneously diagonalizable).

Throughout the field of physics, a classic situation in which two or more operators share a mode structure is when they exhibit the same symmetries or are invariant under transformations of the symmetry group~\cite{Zee2016, Arovas2022}.
Because $\bC$ only contains information about the geometric structure, the symmetry of the LZMs should be completely determined by the system's geometric symmetry. 
On the other hand, the susceptibility ${\partial\vbell / \partial\vbf}$ depends on both the system geometry via its dependence of $\bC$ and the local energetics via its dependence on the stiffness and prestress contained in the Hessian $\bH$.
As a result, we should expect the symmetry of the LZEFs to depend on contributions from both, i.e., the symmetry of the system's overall elastic response.
Using group representation theory~\cite{Zee2016, Arovas2022}, we can formulate these observations into precise statements. 
First, we define what it means for a system to exhibit geometric or energetic symmetry.

We say that a system's geometry exhibits the symmetry of a group $G$ if there exist two unitary matrix representations, ${\bDu_G:G\rightarrow  \mathrm{GL}(\mathbb{R}^{N_{\mathrm{dof}}})}$ and  ${\bDell_G:G\rightarrow \mathrm{GL}(\mathbb{R}^{N_{\mathrm{c}}})}$, that act on node space and bond space, respectively [$\mathrm{GL}(\mathcal{V})$ is the general linear group of the vector space $\mathcal{V}$], such that the bond lengths and node displacements obey the equivariance relation
\begin{equation}
\bDell_G(g)\delta \vbell(\vbu) = \delta \vbell\qty(\bDu_G(g)\vbu)\qc \forall g \in G.\label{eq:invarell}
\end{equation}

Similarly, we say that a system's energy exhibits the symmetry of a group $H$ if there exists a unitary matrix representation, ${\bDell_H:H\rightarrow \mathrm{GL}(\mathbb{R}^{N_{\mathrm{c}}})}$,  that acts on bond space such that the energy obeys the invariance relation
\begin{equation}
E(\vbell(\vbu)) = E\qty(\vbell_0 + \bDell_H(h)\delta \vbell(\vbu))\qc \forall h \in H.\label{eq:invarE}
\end{equation}
For convenience, we construct this representation so that its matrices are identical to those for the geometric symmetry group $G$ for shared group elements, ${\bDell_H(g) = \bDell_G(g)}$ for all $g\in G\cap H$.

Using these definitions, we now formulate the symmetry properties of LZMs and LZEFs into the following theorem (proof provided in Appendix~\ref{app:proofs}):
\begin{theorem}\label{thm:symmetry}
Consider a discrete elastic system described by the Hamiltonian in Eq.~\eqref{eq:hamiltonian}. 
Suppose the geometry exhibits the symmetry of a group $G$ [Eq.~\eqref{eq:invarell}], the energy exhibits the symmetry of a group $H$ [Eq.~\eqref{eq:invarE}], and the elastic response obeys the symmetry of the group $G\cap H$, all with the corresponding matrix representations defined above. Then the following is true:
\begin{enumerate}[(a)]
\item The LZMs obey the symmetry of the geometric symmetry group $G$, forming an invariant subspace under the action of $\bDu_G(g)$ such that for every LZM $\vbuLZM$ and $g\in G$, $\bDu_G(g)\vbuLZM$ is also a LZM.
\item The LZEFs obey the symmetry of elastic response symmetry group $G\cap H$, forming an invariant subspace under the action of $\bDu_{G\cap H}(g)$ such that for every LZEF $\vbfLZEF$ and $g\in G \cap H$, $\bDu_{G\cap H}(g)\vbfLZEF$ is also a LZEF.
\end{enumerate}
\end{theorem}

In the linear regime, Eqs.~\eqref{eq:invarell} and \eqref{eq:invarE} can be used to derive invariance relations for the various matrix operators we have encountered throughout this paper (see Appendix for derivations and a complete list of these invariance relations).
For instance, by taking a derivative of Eq.~\eqref{eq:invarell} with respect to $\vbu$ and evaluated at $\vbu = 0$, we find an invariance relation for the compatibility matrix,
\begin{equation}
\bC = \bDell_G(g^{-1}) \bC \bDu_G(g).\label{eq:invarC}
\end{equation}
Because LZMs are defined as modes in the kernel of $\bC$, it is clear that they should obey the symmetry of the geometry.

Meanwhile, Eq.~\eqref{eq:invarE} implies symmetry of the local energetic properties. Taking derivatives with respect to $\vbell$ and evaluating at $\vbu = 0$, we find
\begin{align}
\vbtp &= \bDell_H(h)\vbtp\label{eq:invartint}\\
\bK &= \bDell_H(h^{-1}) \bK \bDell_H(h)\label{eq:invarK}.
\end{align}

Utilizing both Eqs.~\eqref{eq:invarell} and \eqref{eq:invarE}, we can also take derivatives of the energy with respect to $\vbu$ at $\vbu = 0$ to find that the Hessian transforms as
\begin{equation}
\bH = \bDu_{G\cap H}(g^{-1})\bH\bDu_{G\cap H}(g),\label{eq:invarH}
\end{equation}
with analogous relations for the susceptibilities, including
\begin{equation}
\pdv{\vbell}{\vbf} = \bDell_{G\cap H}(g^{-1}) \pdv{\vbell}{\vbf} \bDu_{G\cap H}(g).\label{eq:invardldf}
\end{equation}
Because LZEFs are defined as modes in the kernel of ${\partial\vbell / \partial\vbf}$, they obey the symmetry of the overall elastic response. 

In short, LZMs are governed by geometric symmetry, while LZEFs are governed by the overall symmetry of the elastic response, which takes into account both the symmetries of the geometry and energetic properties. 
Clearly, if the geometric symmetry group $G$ is a subset of the energetic symmetry group $H$, then the elastic response symmetry group is the same as that of the geometry, $G\cap H = G$, so that the LZMs and LZEFs obey the same symmetries.
However, as we discuss in the next section, this condition is necessary but not sufficient for the LZMs and LZEFs to match.

\subsection{Controlling collective modes via irreducible representations (IRREPs)}

While Theorem~\ref{thm:symmetry} specifies which aspects of a system's symmetry control the symmetries of the LZMs and LZEFs, 
it does not provide enough information to determine whether the modes match or not.
The condition that a set of related LZMs and LZEFs transform under the same symmetry group is necessary, but not sufficient for the modes to become eigenmodes of the Hessian under Corollary~\ref{cor:eigenmode}.
Therefore, our goal is to establish how a system's symmetries constrain the eigenmode structure of the Hessian $\bH$ and the correspondence of these eigenmodes to LZMs and LZEFs. 

To do this, we draw upon some classic results from group representation theory, and in particular, the idea of irreducible representations, or IRREPs~\cite{Zee2016, Arovas2022} (see Appendix~\ref{app:irreps} for a detailed analysis using IRREPs to derive the equations in this section).
Every finite group has a unique set of IRREPs from which all other matrix representations can be constructed. 
Each IRREP (which we denote with the label $\Gamma$) of a group $G$ is a matrix representation of the form ${\bD^\Gamma: G \rightarrow  \mathrm{GL}(\mathcal{V}^\Gamma)}$ where $\mathcal{V}^\Gamma$ is a vector space of dimension $d_\Gamma$ that contains no nontrivial invariant subspaces.
By examining how a general matrix representation breaks down into IRREPs, 
we are provided a natural means to discern the structure symmetry imposes on the vector space upon which it acts.
Any operator that obeys the group symmetries and transforms according to this representation will then contain modes that conform to this structure.

In our case, we are interested in how our node space representation for the elastic response $\bDu_{G\cap H}$ constrains the modes of the Hessian $\bH$ which transforms according to Eq.~\eqref{eq:invarH}.
In particular, one may show that the modes of $\bH$ will organize into multiplets of modes that transform according to different IRREPs upon action by $\bDu_{G\cap H}$. 
To make this explicit, we may construct a complete orthonormal basis for node-space $\ket{\Gamma_s, i}$, where $\Gamma$ is the IRREP under which the modes transform and $i=1,\ldots,d_\Gamma$ is the mode number in a particular multiplet. 
Under the action of $\bDu_{G\cap H}$, each multiplet of modes forms an invariant subspace, transforming as 
\begin{equation}
\bDu_{G\cap H}(g)\ket{\Gamma_s, i} = \sum_{j=1}^{d_\Gamma} \ket{\Gamma_s, j} D^\Gamma_{ji}(g).\label{eq:Dinvarspaces}
\end{equation}
We use the index $s=1,\ldots,n_\Gamma$ to indicate that there may be $n_\Gamma$ mutually orthogonal sets of modes, forming separate invariant subspaces.
In such a basis, our node-space representation $\bDu_{G\cap H}$ will decompose into a block-diagonal form,
\begin{equation}
\bDu{G\cap H}(g) = \sum_\Gamma\sum_{s=1}^{n_\Gamma} \sum_{i,j=1}^{d_\Gamma}\ket{\Gamma_s, i}D_{ij}^\Gamma(g) \bra{\Gamma_s, j},\label{eq:Dblockdiag}
\end{equation}
where each IRREP $\bD^\Gamma$ appears as $n_\Gamma$ repeated block matrices along the diagonal, each of size $d_\Gamma\times d_\Gamma$.

The basis $\ket{\Gamma_s, i}$ that we define above is only partially unique. While the space spanned by all modes with the same IRREP $\Gamma$ and mode number $i$ is uniquely determined by the group symmetries, the way in which the modes break up into different IRREP copies $s$ is not constrained by the symmetry group.
To fully constrain the form of this basis, we must appeal to the details of whatever operator we are interested in.
For any operator that obeys the symmetries of the group $G$, we may choose to orthogonalize the modes so that the operator is fully diagonal. 
In this case, we choose our basis such that the Hessian takes the form
\begin{equation}
\bH = \sum_\Gamma\sum_{s=1}^{n_\gamma}\omega^2_{\Gamma_s}\sum_{j=1}^{d_\Gamma} \ketbra{\Gamma_s, j}.\label{eq:Hblockdiag}
\end{equation}
Here, we see that each copy of each IRREP $\Gamma_s$ is associated with a set of $d_\Gamma$ degenerate energy levels with energy $\omega^2_{\Gamma_s}$ (represented as vibrational frequencies).

With these results in hand, we now understand how a system's symmetries can determine whether the LZMs and LZEFs are the same. 
First, we perform (singular value) decompositions of $\bC$ and ${\partial \vbell / \partial \vbf}$ analogous to Eq.~\eqref{eq:Hblockdiag} and sort the modes into degenerate multiplets according to the IRREP of the appropriate group, $G$ or $G\cap H$, respectively.
Since LZMs and LZEFs are kernel modes, they are guaranteed to be degenerate with singular values of zero.
If both a system's geometry and overall elastic response obey the symmetries of a group,
then we know that the two sets of modes both belong to invariant subspaces of node space.
One way for these subspaces to be the same is if they both correspond to modes that transform according to the same IRREP [Eq.~\eqref{eq:Dinvarspaces}].
However, the subspaces spanned by modes corresponding to different copies of the same IRREP may differ between operators. 
Therefore, we also require that the LZMs and LZEFs span the entire invariant subspace associated with all copies of a given IRREP.
This way, we are guaranteed that the LZMS and LZEFs span the same space. 
We formalize this statement into the following Corollary to Theorem~\ref{thm:symmetry}:
\begin{corollary}\label{cor:subspace}
Consider a discrete elastic system that satisfies Theorem~\ref{thm:symmetry} with a set of LZMs and LZEFs related by Eq.~\eqref{eq:isomorphism},
spanning the vector subspaces $\mathcal{V}^{\mathrm{LZM}}$ and $\mathcal{V}^{\mathrm{LZEF}}$, respectively.
Define the subspace $\mathcal{V}^\Gamma$ spanned by all vectors that transform according to some IRREP $\Gamma$ under the action of the elastic response node space representation $\bDu_{G\cap H}$ [satisfying Eq.~\eqref{eq:Dinvarspaces}],
\begin{equation}
\mathcal{V}^\Gamma = \qty{ \sum_{s=1}^{n_\Gamma}\sum_{i=1}^{d_\Gamma} a_{\Gamma_s, i}\ket{\Gamma_s, i} \middle| a_{\Gamma_s, i} \in \mathbb{R}},
\end{equation}
where $\ket{\Gamma_s, i}$ is an arbitrary complete orthonormal basis that block-diagonalizes $\bDu_{G\cap H}$ [Eq.~\eqref{eq:Dblockdiag}].

If either the LZM subspace $\mathcal{V}^{\mathrm{LZM}}$ or LZEF subspace $\mathcal{V}^{\mathrm{LZEF}}$ is equal to a subspace $\mathcal{V}^\Gamma$ for some IRREP $\Gamma$, then the LZM sand LZEF subspaces are equal, ${\mathcal{V}^{\mathrm{LZM}} = \mathcal{V}^{\mathrm{LZEF}}}$.
\end{corollary}
To see how this Corollary relates Corollary~\ref{cor:eigenmode},  we may choose the basis $\ket{\Gamma_s, i}$ to simply be the eigenbasis of the Hessian as in Eq.~\eqref{eq:Hblockdiag}.
If either the LZMs or LZEFs span the entire invariant subspace spanned by the set of eigenmodes that transform according to IRREP $\Gamma$, then due to their degeneracy, we may write them as eigenmodes of the Hessian, 
thus guaranteeing LZM-LZEF equivalence according to Corollary~\ref{cor:eigenmode}.

\begin{figure}[t!]
\centering
\includegraphics[width=\linewidth]{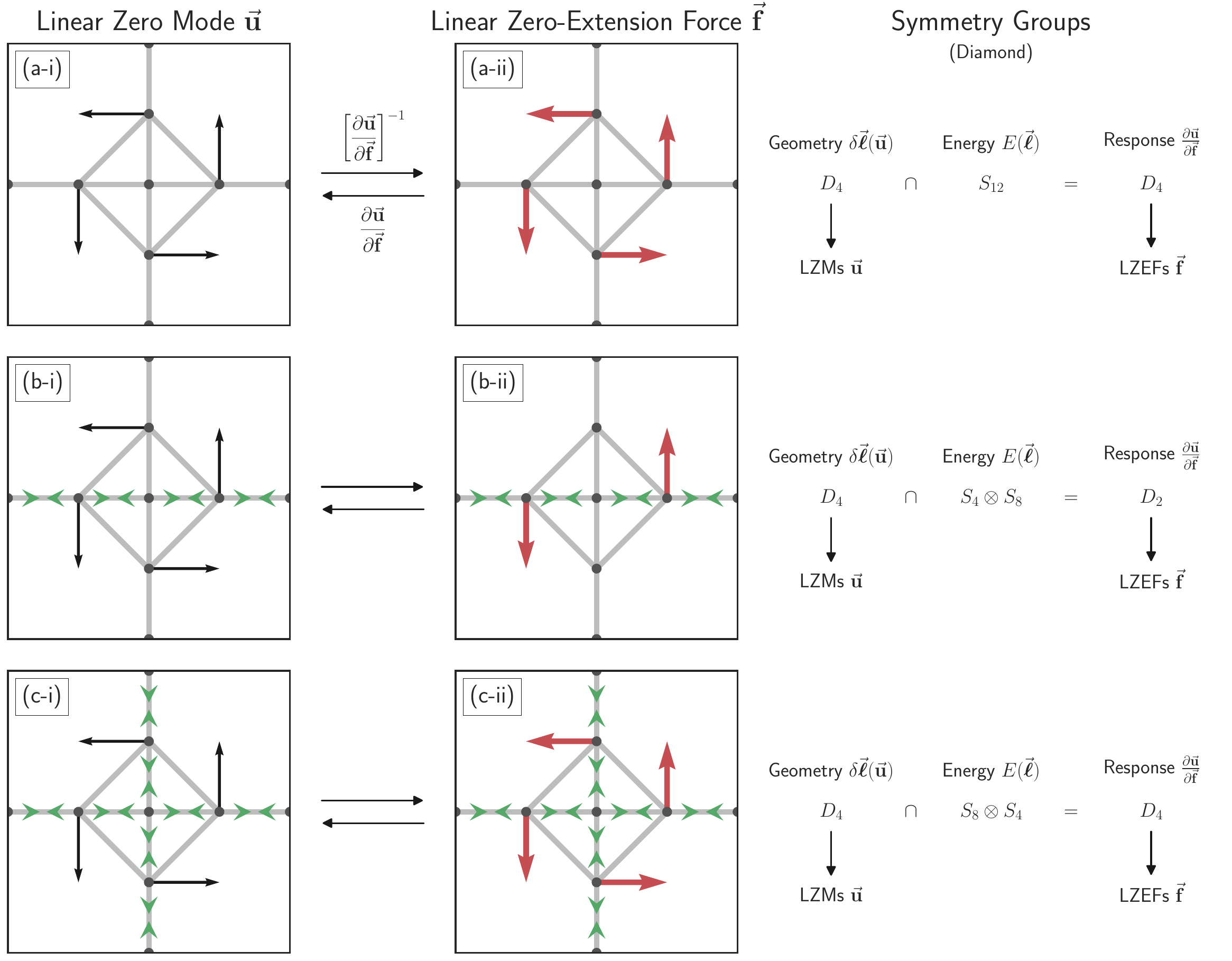}
\caption{\textbf{Breaking and restoring dihedral symmetry with prestress.}
A diamond-shaped central-force spring network with fixed boundary conditions whose geometry exhibits four-fold dihedral symmetry $D_4$.
In each row, a linear zero mode (LZM) and the linear zero-extension force (LZEF) that couples to it  via the susceptibility ${\partial \vbu / \partial\vbf}$ are shown for a different choice of prestress.
For each case, the rightmost column lists the symmetry groups describing the geometry, energy (prestress), elastic response, LZM, and LZEF.
In all cases, the spring constants are chosen to be uniform ($\bK = \bI$).
\textbf{(a)}~When the prestress is zero ($\vbtp = 0$) the energy exhibits full permutation symmetry $S_{12}$ which encompasses the symmetry of the geometry, resulting in overall $D_4$ symmetry. The (a-i) LZM and its corresponding (a-ii) LZEF  are identical, both exhibiting full four-fold dihedral symmetry.
\textbf{(b)}~Introducing nonzero prestress along the central line of horizontal bonds breaks the energetic symmetry to $S_4\otimes S_8$, reducing overall symmetry of the response to two-fold dihedral symmetry $D_2$.
While the (b-i)  LZM  is the same, the (b-ii)  LZEF that couples to it no longer matches.
\textbf{(c)}~Introducing additional prestress along the central line of vertical bonds restores some of the energetic symmetry to $S_4\otimes S_8$ so that the overall response exhibits four-fold dihedral symmetry $D_4$.
The  (c-i) LZM  and the (c-ii)  LZM  that couples to it match once again.}
\label{fig:symbreak}
\end{figure}

\begin{figure*}[t!]
\centering
\includegraphics[width=\linewidth]{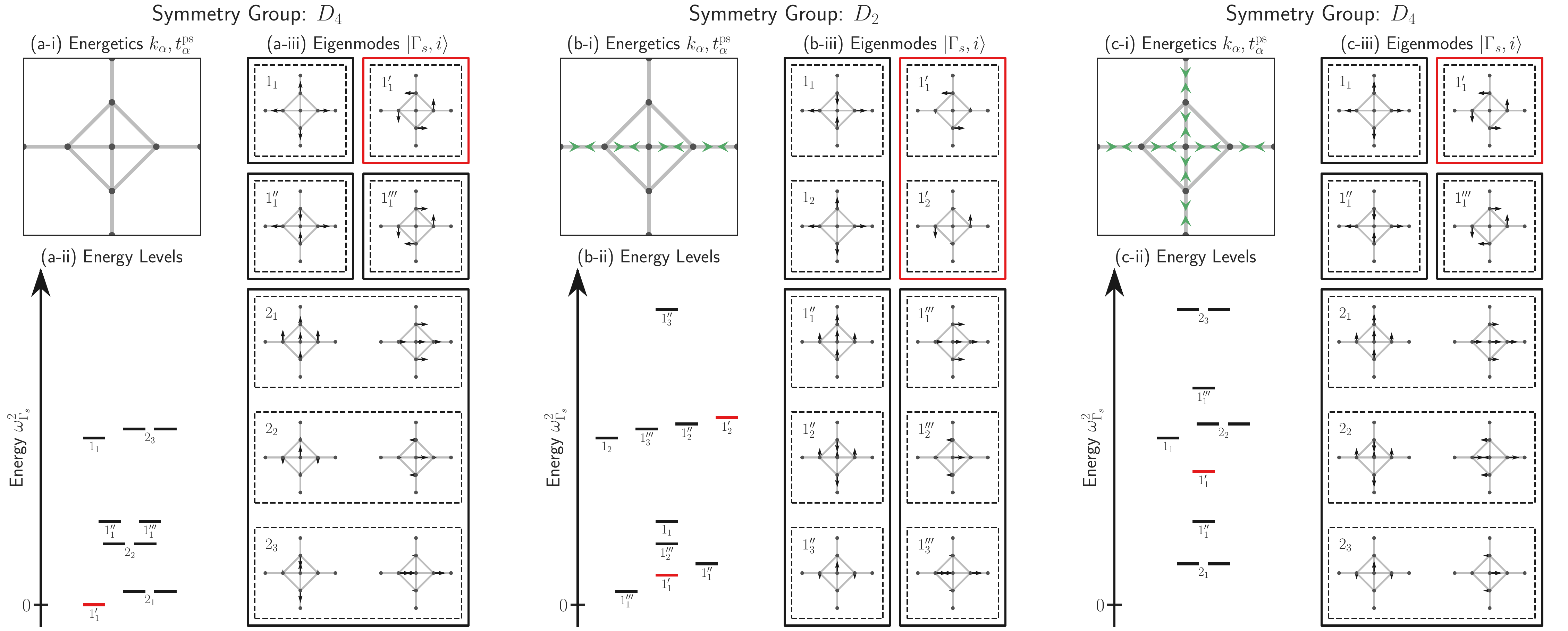}
\caption{
\textbf{Eigenmode structure and irreducible representations of diamond network.}
The energy levels and eigenmodes for each configuration in Fig.~\ref{fig:symbreak} labeled according to IRREP.
In each column, the local energetic properties are depicted in the upper-left panel. The bond stiffnesses $k_\alpha$ are chosen to be uniform for each network, while the prestress $\tp_\alpha$ is shown with green arrows.
In the bottom-right, we show the energy levels $\omega^2_{\Gamma_s}$ (eigenvalues of the Hessian written as squared vibrational frequencies). Each energy is labeled with an IRREP $\Gamma$ and a copy number $s$. 
In the right panel, we depict the eigenmodes of the Hessian $\ket{\Gamma_s, i}$ corresponding to each energy level.
Solid boxes group together modes that transform according to the same IRREP $\Gamma$, while dashed boxes group together degenerate modes corresponding to a single copy $s$ of $\Gamma$.
In red we highlight the energy level and eigenmodes that couple to the corresponding LZMs and LZEFs in Fig.~\ref{fig:symbreak}.
\textbf{(a)} When the symmetry of the elastic response matches the geometric symmetry, the LZM spans the entire one-dimensional invariant subspace corresponding to the $1'$ IRREP of $D_4$.
As a result, the LZM and LZEF are guaranteed to match.
\textbf{(b)} When the energetic symmetry is more restrictive than the geometric symmetry, the the LZM now couples to two different eigenmodes from two different copies of the $1'$ IRREP for $D_2$, but does not span the entire two-dimensional subspace spanned by all copies of $1'$. Therefore, the LZM and LZEF are no longer guaranteed to match.
\textbf{(c)} The symmetry of the elastic response matches the geometric symmetry again, so the LZM again spans the entire invariant subspace corresponding to the $1'$ IRREP of $D_4$. The LZM and LZEF now both match again.
}
\label{fig:levels}
\end{figure*}

\begin{figure*}[t!]
\centering
\includegraphics[width=\linewidth]{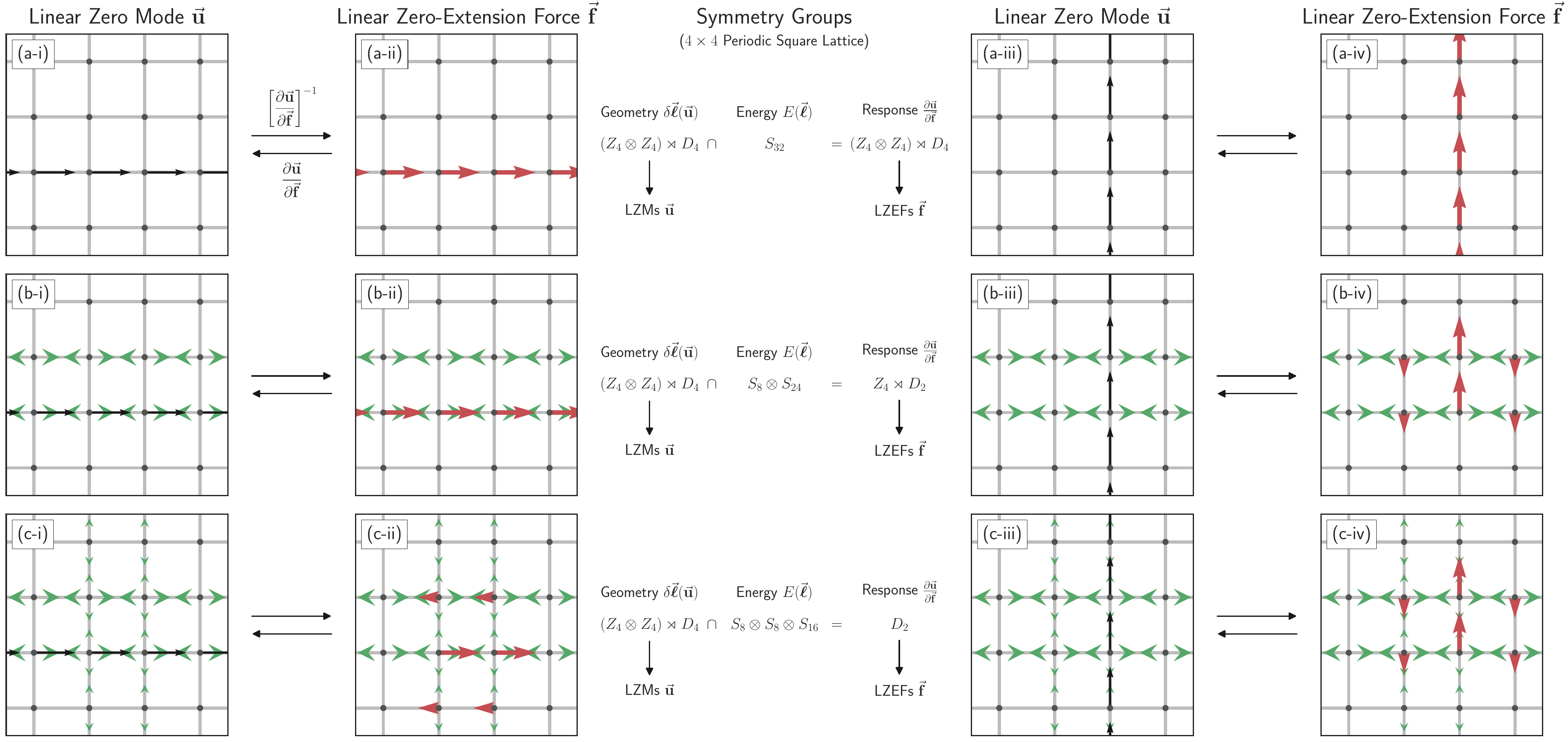}
\caption{\textbf{Repeatedly breaking translational symmetry with prestress.} A central-force spring network forming a $4\times 4$ periodic square lattice.
The network geometry exhibits four-fold dihedral symmetry, along with discrete translational symmetry along the $x$- and $y$-axes, forming the group $(Z_4 \otimes Z_4)\rtimes D_4$.
In each row, two linear zero modes (LZMs) and the linear zero-extension forces (LZEFs) that couple to them via the susceptibility ${\partial \vbu / \partial\vbf}$ are shown for a different choice of prestress.
The central column lists the symmetry groups for each case describing the geometry, energy (prestress), elastic response, LZMs, and LZEFs.
In all cases, the spring constants are chosen to be uniform ($\bK = \bI$).
\textbf{(a)}~For zero prestress ($\vbtp = 0$) the energy exhibits full permutation symmetry $S_{32}$, resulting in overall $(Z_4 \otimes Z_4)\rtimes D_4$ symmetry. 
Each LZM matches the LZEF that couples to it, with one pair, (a-i) and (a-ii), exhibiting $x$-translational symmetry, and the other pair, (a-iii) and (a-iv), exhibiting $y$-translational symmetry.
\textbf{(b)}~Introducing nonzero prestress along two horizontal lines of bonds ($\tp_\alpha = 0.8$) breaks the $y$-translational, symmetry reducing the energetic symmetry to $S_8\otimes S_{24}$, and the overall elastic response to $Z_4\rtimes D_2$.
The (b-iv) LZEF  that previously displayed $y$-translational symmetry no longer matches its (b-iii) LZM , while the  (b-ii) LZEF that displayed $x$-translational symmetry is unaffected.
\textbf{(c)}~Introducing additional prestress ($\tp_\alpha = 0.4$) along two vertical lines of bonds, but with smaller magnitude than the prestressed horizontal bonds, breaks $x$-translational symmetry, reducing the energetic symmetry to ${S_8\otimes S_8\otimes S_{24}}$, and the overall elastic response to two-fold dihedral symmetry $D_2$.
In this case, neither LZEF matches it corresponding LZM.
}

\label{fig:subgroup}
\end{figure*}

\begin{figure}[t!]
\centering
\includegraphics[width=\linewidth]{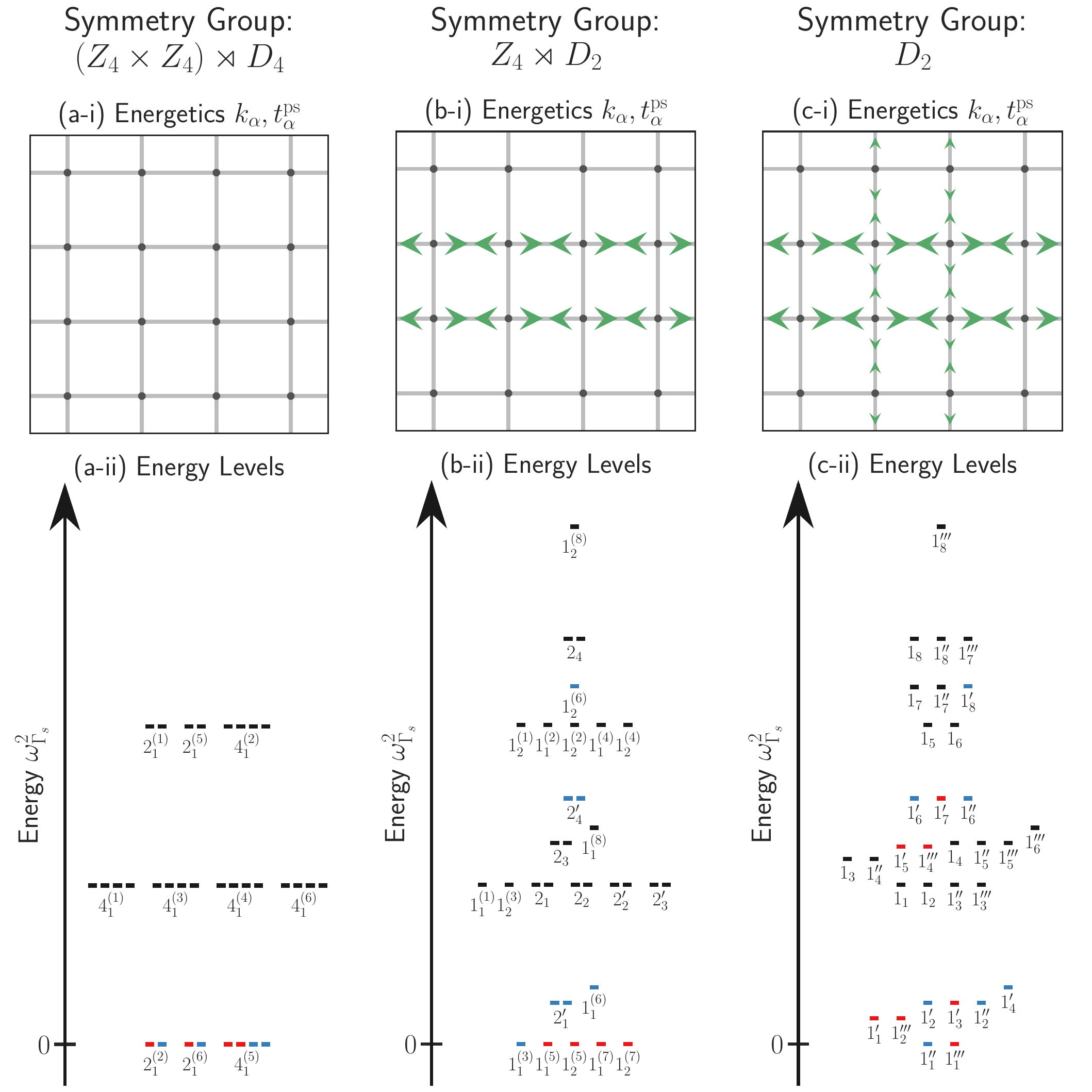}
\caption{
\textbf{Energy levels and irreducible representations of $4\times 4$ square lattice.}
The energy levels for each configuration in Fig.~\ref{fig:subgroup} labeled according to IRREP.
The local energetic properties are depicted in the top row. The bond stiffnesses $k_\alpha$ are chosen to be uniform for each network, while the prestress $\tp_\alpha$ is shown with green arrows.
In the bottom row, we show the energy levels $\omega^2_{\Gamma_s}$, each with an IRREP and copy number $s$.
In red and blue we highlight the energy levels that couple to the LZMs with $x$- and $y$-translational symmetry, respectively [see Fig.~\ref{fig:subgroup}].
\textbf{(a)} When the elastic response symmetry matches the geometric symmetry, the LZMs span the entire invariant subspaces corresponding to the $2^{(2)}$, $2^{(6)}$, and $4^{(5)}$ IRREPs of $(Z_4\otimes Z_4)\rtimes D_4$.
As a result, the LZMs and LZEFs are guaranteed to match.
\textbf{(b)} Upon breaking $y$-translational symmetry with prestress, the LZMs with $y$-translational symmetry no longer span the entire invariant subspace for any IRREP and no longer match their LZEFs.
In contrast, preservation of $x$-translational symmetry causes the LZMs with $x$-translational symmetry to fully span the subspaces corresponding to the  the $1^{(5)}$ and $1^{(7)}$ IRREPs of $Z_4\rtimes D_2$ and therefore match their LZEFs.
\textbf{(c)} Upon further breaking $x$-translational symmetry with prestress, the LZMs with $x$-translational symmetry also no longer span the entire invariant subspace for any IRREP, so that none of the LZMs are guaranteed to match their LZEFs.
}
\label{fig:levels4x4}
\end{figure}

\subsection{Example: Breaking and restoring symmetry}

In Figs.~\ref{fig:symbreak} and \ref{fig:levels} we use a simple diamond-shaped spring network with fixed boundary conditions to demonstrate how Theorem~\ref{thm:symmetry} and Corollary~\ref{cor:subspace}  may be used to control the relationship between LZMs and the LZEFs that couple to them.
The nodes and bonds of this network exhibit four-fold rotational symmetry with four axes of reflection symmetry about the center (vertical and horizontal axes, plus two diagonal axes).
In the language of group theory, the symmetry of this network is captured by the dihedral group $D_4$.
The network geometry is invariant under transformations of this group, so according to Theorem~\ref{thm:symmetry}, we expect the LZMs to also exhibit the same $D_4$ symmetry.
This network has a single LZM, shown in Fig.~\ref{fig:symbreak}(a-i), in which the central diamond rigidly rotates anticlockwise.
In accordance with Theorem~\ref{thm:symmetry}, this mode forms an invariant subspace under $D_4$; it is preserved under rotations by $90^\circ$ and only changes up to a sign under the allowed reflections about the center.

To understand why this set of displacements is a LZM, we remark that for central-force springs, 
the extension of a bond $\alpha$ connecting nodes $i$ and $j$ can be expressed to to linear order as ${\delta \ell_\alpha = \hat{b}_{ij}\cdot (\vec{u}_j-\vec{u}_i)}$ where $\vec{u}_i$ and $\vec{u}_j$ are $d$-dimensional vectors of displacements for each node and $\hat{b}_{ij}$ is a unit vector pointing from $j$ to $i$.
This means that the compatibility matrix $\bC$ only captures relative displacements of the nodes parallel to the bonds.
In contrast, motion perpendicular to the bonds does not result in any bond extensions to linear order,
as is the case in Fig.~\ref{fig:symbreak}(a).
This property is well known in the literature~\cite{Lubensky2015}.

Theorem~\ref{thm:symmetry} states that the symmetry of the LZEF that couples to this LZM should be determined by the overall elastic response which depends on both the geometry and energy.
To test this, we first choose all spring constants to be identical, $\bK = \bI$, and set the prestress to zero, $\vbtp = 0$.
In this case, the local energetics trivially exhibit twelve-fold permutation symmetry described by the permutation group $S_{12}$ and are invariant according to both Eqs.~\eqref{eq:invarK} and \eqref{eq:invartint}.
It is simple to convince oneself that the geometric symmetry group $D_4$ is a subset of the more general energetic symmetry group $S_{12}$ so that the symmetry of the total elastic response captured by their intersection is also $D_4$.
Next, we apply the inverse of the susceptibility ${\partial \vbu / \partial\vbf}$ to map the LZM in Fig.~\ref{fig:symbreak}(a-i) to its corresponding LZEF, depicted in Fig.~\ref{fig:symbreak}(a-ii).
As expected, we find that the LZEF also exhibits $D_4$ symmetry and forms an invariant subspace, matching the LZM.

To understand why the LZM and LZEF match, we examine the IRREP structure of the Hessian for this network under $D_4$ (see Appendix~\ref{app:diamond} for node space representation and IRREP matrices). 
In Figs.~\ref{fig:levels}(a-ii) and (a-iii), we depict the energy levels $\omega^2_{\Gamma_s}$ (eigenvalues of the Hessian written as squared vibrational frequencies) and eigenmodes $\ket{\Gamma_s, i}$ for this network, respectively.
Each energy level and eigenmode is labeled according to the IRREP $\Gamma_s$ of $D_4$ under which they transform [Eq.~\eqref{eq:Dinvarspaces}]. 
The subscript $s$ indicates the copy of IRREP $\Gamma$ in the node space representation.
We see that the eigenmodes break up into four one-dimensional invariant subspaces that transform according to IRREPs $\Gamma = 1$, $1'$, $1''$, or $1'''$, and three two-dimensional invariant subspaces that each transform according to a different copy of IRREP $2$, each with a pair of degenerate eigenvalues (we also observe an accidental degeneracy between IRREPs $1''$ and $1'''$ that can be broken by applying nonzero prestress).
In red, we highlight the energy level and mode that couple to the LZM, 
 which happen to correspond to the $1'$ IRREP.
In this case, applying an LZM costs zero energy to lowest order due to the lack of prestress.

Now according to Corollary~\ref{cor:subspace}, if the LZM or LZEF span the entire subspace formed by all copies of an IRREP, they will be guaranteed to match.
Since there is only one copy of the $1'$ IRREP and it is one-dimensional, this condition is satisfied.
Therefore, both modes are guaranteed to be eigenmodes of the Hessian and to match, as we can observe.

Next, we break energetic symmetry by adding nonzero prestress with constant magnitude, $\tp_\alpha = 0.8$, along the horizontal bonds (this preserves the ground state configuration) as shown in Fig.~\ref{fig:symbreak}(b).
Due to this prestress, the symmetry group of the energy is reduced to the product of two smaller permutation groups, $S_4\otimes S_8\subset S_{12}$, with permutations no longer allowed between bonds with and without prestress.
The result is that the overall symmetry of the response is reduced to the two-fold dihedral group $D_2$, the intersection of $D_4$ and $S_4\otimes S_8$.
Because the LZM relies only on geometry and is blind to the energetics, it remains unchanged [Fig.~\ref{fig:symbreak}(b-i)].
In contrast, LZEFs are sensitive to energetic costs,
and we find that the LZEF that couples to this LZM no longer points in the same direction, but instead localizes to the line of prestressed bonds.
Consistent with the symmetry of the elastic response, the LZEF now exhibits two-fold dihedral symmetry $D_2$ [Fig.~\ref{fig:symbreak}(b-ii)].

Examining the eigenmode structure of the network in Figs.~\ref{fig:levels}(b-ii) and (b-iii), we see that the LZM now couples to two different modes that transform according to two different copies of the $1'$ IRREP of $D_2$ (highlighted in red).
However, the single LZM in the system cannot span the entire two-dimensional subspace belonging to both copies of $1'$.
As a result, Corollary~\ref{cor:subspace} no longer provides a guarantee that the LZM will match its LZEF.

Finally, in Fig.~\ref{fig:symbreak}(c), we restore some symmetry by introducing additional prestress along the vertical bonds with the same magnitude as the horizontal ones, resulting in the energetic symmetry group $S_8\otimes S_4$.
Now, the geometric $D_4$ symmetry is once again a subset of the larger energetic symmetry group, resulting in an overall $D_4$ symmetry for the elastic response.
Consequently, we find that the LZEF  [Fig.~\ref{fig:symbreak}(c-ii)] and LZM [Fig.~\ref{fig:symbreak}(c-i)] both exhibit $D_4$ symmetry in accordance with Theorem~\ref{thm:symmetry}.
In Figs.~\ref{fig:levels}(c-ii) and (c-iii), we also see that once again, the LZM couples to a single eigenmode of the Hessian which spans the subspace that transforms with the $1'$ IRREP of $D_4$,
thus guaranteeing that it matches its LZEF in according to Corollary~\ref{cor:subspace}.

In the rightmost column of Fig.~\ref{fig:symbreak}, we summarize these symmetry results by specifying the symmetry groups of the system's geometry, energy, and elastic response, along with those of the depicted LZMs and LZEFs. 
In Appendix~\ref{app:diamond}, we provide formal group definitions, node space representation matrices, and IRREP matrices.

\subsection{Example: Repeated symmetry breaking}

The symmetry breaking shown in the previous example can be taken one step further.
We can use prestress to successively break the symmetry group describing a system's elastic response to more and more restrictive subgroups.
By introducing prestress that preserves some, but not all, of a system's geometric symmetries,
we can control the resulting level of symmetry exhibited by the LZEFs due to the elastic response.

We demonstrate this behavior in Fig.~\ref{fig:subgroup} for a $4\times 4$ square lattice of springs with periodic boundary conditions.
As before, this network's bonds and nodes exhibit four-fold dihedral symmetry described by the group $D_4$.
The $4\times 4$ square grid also adds discrete translational symmetry along the $x$- and $y$-axes.
Because translations along the two axes commute, the overall translational symmetry is described by the product group $Z_4\otimes Z_4$.
To obtain the overall geometric symmetry group, we take the semidirect product of the two groups to obtain $(Z_4\otimes Z_4)\rtimes D_4$ (the semidirect product arises because $Z_4\otimes Z_4$ is a normal subgroup, while $D_4$ is not).

This network contains a total of eight LZMs, each consisting of nodes moving along one of the eight horizontal or vertical filaments of bonds.
In Figs.~\ref{fig:subgroup}(a-i) and \ref{fig:subgroup}(a-iii), we show a pair of these LZMs, exhibiting $x$- and $y$- translational symmetry, respectively.
By applying transformations of the network's geometric symmetry group to either one of these modes, we can generate all eight of the network's LZMs.

As in the previous example, we first consider the case where the spring constants are identical, $\bK = \bI$, and set the prestress to zero, $\vbtp = 0$, resulting in full $S_{32}$ permutation symmetry of the bond stiffnesses and prestresses.
Since the geometric symmetry is a subset of the energetic symmetry, the symmetry of the overall elastic response is the same as the geometry, $(Z_4\otimes Z_4)\rtimes D_4$.
As a result, the LZEFs exhibit the same symmetries as the LZMs. 
As shown in Figs.~\ref{fig:subgroup}(a-ii) and \ref{fig:subgroup}(a-iv), we find that each LZM couples to an identical LZEF.

In Fig.~\ref{fig:levels4x4}(a-ii), we depict the energy levels for this network and classify them according to the IRREPs of $(Z_4\otimes Z_4)\rtimes D_4$ (see Appendix~\ref{app:lattice} for node space representation and IRREP matrices).
In red and blue, we highlight the energy levels of the modes that couple to the LZMs consisting of motion along the $x$- and $y$-directions, respectively, exhibiting $x$- and $y$-translational symmetry. 
We see that the eight LZMs fully span all eight modes corresponding to all copies of the IRREPs $2^{(2)}$, $2^{(6)}$, and $4^{(5)}$.
From Corollary~\ref{cor:subspace}, we conclude that the LZMs are guaranteed to match the LZMs, consistent with our expectations based on Figs.~\ref{fig:subgroup}.

Next, in Fig.~\ref{fig:subgroup}(b), we break the system's discrete translational symmetry along the $y$-axis by introducing prestress along two horizontal rows of bonds of size $\tp_\alpha = 0.8$.
While both LZMs are preserved [Figs~\ref{fig:subgroup}(b-i) and \ref{fig:subgroup}(b-iii)], only the LZEF exhibiting $x$-translational symmetry remains the same [Fig.~\ref{fig:subgroup}(b-ii)].
In contrast, the other LZEF [Fig.~\ref{fig:subgroup}(b-iv)] no longer exhibits $y$-translational symmetry and localizes to the two rows of prestressed bonds.
The symmetry of the energy has broken to the subgroup ${S_8\otimes S_{24} \subset S_{32}}$. As a result, the symmetry of the elastic response is reduced to $Z_4 \rtimes D_2$,
consisting of translation symmetry along the $x$-axis and two-fold dihedral symmetry.

Examining the energy levels in Fig.~\ref{fig:levels4x4}(b-ii) classified according to the IRREPs of $Z_4 \rtimes D_2$, we find that the four LZMs with $x$-translational symmetry fully span all four modes corresponding to all copies of the IRREPs $1^{(5)}$ and $1^{(7)}$ (highlighted in red). 
This is consistent with our expectations that the LZEFs for these LZMs should match because the system still exhibits overall $x$-translational symmetry. 
In contrast, we now observe that the LZMs with $y$-translational symmetry do not completely span all modes belonging to all the copies of any IRREP, instead coupling to a total of seven different modes from three different IRREPs.
As a result, their LZEFs are no longer guaranteed to match.

Finally, we break the network's $x$-translational symmetry as well.
In Fig.~\ref{fig:subgroup}(c), we introduce additional prestress along two vertical lines of bonds, but with magnitude less than that of the prestressed horizontal bonds of size $\tp_\alpha = 0.4$.
We now find that neither of the LZEFs [Figs.~\ref{fig:subgroup}(c-ii) and \ref{fig:subgroup}(c-iv)]  matches its corresponding LZM [Figs.~\ref{fig:subgroup}(c-i) and \ref{fig:subgroup}(c-iii)].
The symmetry of the energy has broken further to the subgroup ${S_8\otimes S_8 \otimes S_{16} \subset S_8\otimes S_{24}}$, resulting in an overall two-fold dihedral symmetry $D_2$ for the elastic response.

A look at the energy levels in Fig.~\ref{fig:levels4x4}(b-iii) classified according to the IRREPs of $D_2$ reveals that the LZMs with $x$-translational symmetry also no longer fully span the modes corresponding to any IRREPs, coupling to seven different modes just like the LZMs with $y$-translational symmetry.
As a result, none of the LZEFs are guaranteed to match their LZMs.

In the central column of Fig.~\ref{fig:subgroup}, we summarize these symmetry results by specifying the symmetry groups of the system's geometry, energy, and elastic response, along with those of the depicted LZMs and LZEFs. 
In Appendix~\ref{app:lattice}, we provide the group definitions, node space representations, and IRREP matrices for groups describing the symmetry of the elastic response used to create Figs.~\ref{fig:subgroup} and \ref{fig:levels4x4}.

\section{Discussion}\label{sec:discussion}

In summary, we used an approach based on susceptibilities to derive a generalization of the Maxwell-Calladine index theorem that includes local energetic properties like stiffness and prestress.
In this generalized version, the classical linear zero modes (LZMs) and states of self-stress (SSSs) are replaced with linear zero-extension forces (LZEFs) and zero-displacement tensions (ZDTs).
While the LZMs and SSSs derive purely from the linearized geometric relationships between constraints and degrees of freedom,
LZEFs and ZDTs are types of external perturbations that necessarily depend on both a system's geometric structure and the details of the local energetic deformation costs.
We then explored the detailed relationship between LZEFs and ZDTs and their classical counterparts.
While ZDTs and SSSs are identical, LZEFs and LZMs are generally related by a nontrivial isomorphism, with each LZEF coupling to a unique LZM.

Controlling the relationships between LZEFs and LZMs represents an interesting avenue for designing mechanical metamaterials with specific responses to external forces.
As we demonstrated using group representation theory, whether or not LZEFs couple to identical LZMs depends on whether the group describing the symmetry of a system's elastic energy encompasses that of its structural geometry.
Specifying the symmetry of the local energy (e.g., via prestress) can then be used to control the forms of the LZEFs that couple to the LZMs.

Our analysis and examples presented in this paper focused on LZMs and LZEFs, modes that preserve the lengths of bonds to linear order.
However, we would expect the design principles we have demonstrated should apply to all types of displacements and forces.
In general, all eigenmodes of the Hessian, whether or not they are LZMs or LZEFs, should transform under the symmetries of the elastic response, which can be controlled via bond stiffness, prestress, and geometry.
In fact, the proofs of Theorem~\ref{thm:symmetry} and Corollary~\ref{cor:subspace} do not rely on the fact that a set of displacements are LZMs, but rather only on the property that they are degenerate modes of compatibility matrix.

In this paper, susceptibilities to external perturbations provided a natural language for exploring mechanical rigidity and the relationship between constraints and degrees of freedom.
This framework provides a natural extension to previous formalism proposed in Ref.~\onlinecite{Connelly2022}, where the authors relate the different vector spaces in a picture similar to Fig.~\ref{fig:maps}.
In that work, the authors label node space and bond space as the external and internal vector spaces, respectively, and connect these two spaces via the compatibility matrix as we do here.
However, they then connect displacements and extensions to forces and tensions via the principle of virtual work, 
whereas we use the explicit maps provided by susceptibilities, 
allowing us to easily transform between any of the four types of vectors.
 
We also believe that an approach based on susceptibilities could provide a path out of the linear regime. 
The linear susceptibilities we compute in Eq.~\eqref{eq:suscept} and the generalized Maxwell-Calladine relation in Eq.~\eqref{eq:genMC} should in principle capture all rigidity information described by the linear response.
However, they still do not fully capture the effects of prestress-induced rigidity. 
This is evident in the examples shown in Figs.~\ref{fig:symbreak}-\ref{fig:levels4x4}, where the numbers of LZMs and LZEFs do not change even if prestress stabilizes the system because the geometry deforms at second order, but not first order.
This makes a strong case that a general theory of prestress stability should require an analysis of the nonlinear regime.
Higher-order susceptibilities such as ${\partial^2 \vbell / \partial \vbf^2}$
could prove useful for deriving conditions for prestress stability beyond linear order such as those in Refs.~\onlinecite{Damavandi2022, Damavandi2022a}.
On a related note, these nonlinear susceptibilities could be used to understand anomalous rigidity in materials that appear underconstrained to linear order such as packings of ellipses~\cite{Donev2007, Mailman2009} or cell vertex models~\cite{Hernandez2022}.
They may also provide a means to integrate energetic properties into a theory of topological protection for nonlinear mechanisms~\cite{Lo2021}.
Alternatively, they may elucidate the scaling behavior of material properties near rigidity phase transitions~\cite{Goodrich2016}, just as they have been used historically to investigate phase transitions throughout the field of statistical physics.

It would also be interesting to combine such a formalism with group representation theory to design materials that undergo specific finite deformations in response to external forces.
Already, group theoretic descriptions have been developed to analyze highly symmetric tensegrity structures to ascertain whether LZMs extend to nonlinear mechanisms and to analyze prestress-stability~\cite{Schulze2010, Connelly2022}, but not to control responses to external forces.

Finally, in our formalism, each discrete component imposes a soft constraint on the degrees of freedom associated with its energetic cost of deformation.
In some systems, it is often useful to also include hard constraints, 
such as in foams where the Plateau rule restricts the angles of the films at each vertex to 120$^\circ$~\cite{Weaire2001}, 
or in origami structures where one may wish to impose fixed areas to the facets~\cite{Schenk2016}.
It would be interesting to extend our formalism to explicitly include such hard constraints and repeat the analysis performed here to discover how it modifies the relationships between collective modes.

\section*{Acknowledgments}

We would like to thank  Daniel Sussman and Alexander Golden for extremely useful discussions. 
This work was supported by NIH NIGMS grant 1R35GM119461 award to P.M.


%

\appendix

\setcounter{table}{0}
\renewcommand{\thetable}{S\arabic{table}}

\setcounter{figure}{0}
\renewcommand{\thefigure}{S\arabic{figure}}

\section{Derivation of Mechanical Susceptibilities}\label{app:derivation}

\subsection{Exact Formulas}

Here, we derive the expressions for the susceptibilities in Eq.~\eqref{eq:suscept}. 
To begin, we observe that in the absence of external forces or tensions, the ground state is determined by force equilibrium, i.e., the internal net forces are zero,
\begin{equation}
f^{\mathrm{int}}_i \equiv \pdv{E}{u_i} = 0,
\end{equation}
where $E$ is the energy defined by Eq.~\eqref{eq:energy}.
For convenience, we use the ground state configuration (which may not always be unique) as the reference state for each degree of freedom so that $\vbu=0$ at the ground state,
\begin{equation}
\pdv{E}{u_i}\eval_{\vbu=0} = 0.
\end{equation}
We also assume we are dealing with a stable ground state.

Next, we take the gradient of the Hamiltonian in Eq.~\eqref{eq:hamiltonian} with respect to the displacements and set it to zero, giving us
\begin{equation}
0 = \pdv{\mathcal{H}}{u_i} = \pdv{E}{u_i} + \lambda u_i - f_i- \sum_\alpha t_\alpha \pdv{\ell_\alpha}{u_i}.
\end{equation}
Taking a second derivative with respect to $f_j$, we find
\begin{equation}
0 = \sum_k\qty[H_{ik}+\lambda\delta_{ik} - \sum_\alpha t_\alpha \pdv{\ell}{u_i}{u_k}]\pdv{u_k}{f_j} - \delta_{ij}.
\end{equation}
where $H_{ik}$ is the Hessian which takes the form in Eq.~\eqref{eq:Hessian}.
When $t_\alpha=0$, this becomes
\begin{equation}
0 = \sum_k\qty[H_{ik}+\lambda\delta_{ik}]\pdv{u_k}{f_j} - \delta_{ij}.
\end{equation}
Changing over to matrix form, this becomes
\begin{equation}
0 = \qty[\bH + \lambda \bI]\pdv{\vbu}{\vbf} - \bI.
\end{equation}
Because the ground state is stable, $\bH$ is guaranteed to be positive semi-definite and this equation can readily be solved, giving us
\begin{equation}
\pdv{\vbu}{\vbf} = \qty[\bH + \lambda \bI]^{-1}.\label{eq:dudfexact}
\end{equation}
Using the chain rule, we can use this solution to find the susceptibility for the extensions with respect to forces,
\begin{equation}
\pdv{\vbell}{\vbf} = \pdv{\vbell}{\vbu}\pdv{\vbu}{\vbf} = \bC \qty[\bH + \lambda \bI]^{-1}.
\end{equation}
Similarly, taking the derivative of the relation between external tensions and forces from Eq.~\eqref{eq:Crelations}, $\vbf = \bC^T\vbt$, we find that the following is true for forces that can be completely expressed in terms of tensions:
\begin{equation}
\pdv{\vbf}{\vbt} = \bC^T.
\end{equation}
Using this, we can derive the remaining two susceptibilities using the chain rule,
\begin{align}
\pdv{\vbu}{\vbt} &= \pdv{\vbu}{\vbf}\pdv{\vbf}{\vbt} =  \qty[\bH + \lambda \bI]^{-1}\bC^T\\
\pdv{\vbell}{\vbt} &= \pdv{\vbell}{\vbf}\pdv{\vbf}{\vbt} =  \bC \qty[\bH + \lambda \bI]^{-1}\bC^T.
\end{align}

In summary, the exact formulas for the four susceptibilities are
\begin{equation}
\mqty(\pdv{\vbu}{\vbf} & \pdv{\vbu}{\vbt}\\
\pdv{\vbell}{\vbf} & \pdv{\vbell}{\vbt}) = 
\mqty( \qty[\bH + \lambda \bI]^{-1} &  \qty[\bH + \lambda \bI]^{-1} \bC^T\\
 \bC\qty[\bH + \lambda \bI]^{-1} &  \bC\qty[\bH + \lambda \bI]^{-1}\bC^T).\label{eq:exact}
\end{equation}

\subsection{Matrix Psuedoinverse}\label{app:pseudoinverse}

Next, we would like to take the limit in which the regularization constant goes to zero, $\lambda \rightarrow 0$.
To do this, we first define the pseudoinverse of a general matrix $\bA$ of size $n\times m$. 
We define the left and right singular vectors of $\bA$ as $\ket{l_i}$ and $\bra{r_i}$ with positive singular values $\sigma_i$, where $i=1,\ldots, k$ with $k\leq m, n$.
In this notation, we can write the singular value decomposition of $\bA$ as
\begin{equation}
\bA = \sum_i \sigma_i \ketbra{l_i}{r_i}.
\end{equation}
We can then define the pseudoinverse of $\bA$ as 
\begin{equation}
\bA^+ = \sum_i \frac{1}{\sigma_i} \ketbra{r_i}{l_i}.
\end{equation}
Note that the singular values become inverted and that the left and right singular vectors switch places.

For a real symmetric positive semi-definite matrix $\bA$, we will use the following identity for small $\lambda$:
\begin{equation}
\qty[\bA + \lambda \bI]^{-1} \approx \frac{1}{\lambda}\qty[\bI - \bA\bA^+] + \sum_{n=0}^\infty (-\lambda)^n(\bA^+)^{n+1}.\label{eq:invexpand}
\end{equation}
To prove this, we note that for a real symmetric matrix, the singular value decomposition and the eigendecomposition are identical because the left and right singular vectors are the same and can be chosen to form an orthonormal basis. We may write the eigendecomposition of $\bA$ as,
\begin{equation}
\bA = \sum_i \sigma_i \ketbra{i},
\end{equation}
where $\ket{i}$ are the eigenvectors and $\sigma_i$ are the eigenvalues which are real and non-negative.
Sorting the eigenvalues into those where  $\sigma_i=0$ and $\sigma_i > 0$,
we express the pseudoinverse as
\begin{equation}
\bA^+ = \sum_{i: \sigma_i > 0}\frac{1}{\sigma_i}\ketbra{i}.
\end{equation}
and the projector onto the kernel of $\bA$ as
\begin{equation}
\bI - \bA\bA^+ = \sum_{i: \sigma_i = 0}\ketbra{i}.
\end{equation}
We can then derive the identity by expanding the inverse in small $\lambda$, 
\begin{equation}
\begin{aligned}
\qty[\bA + \lambda \bI]^{-1} &= \sum_i \frac{1}{\sigma_i+\lambda} \ketbra{i}\\
&\approx \frac{1}{\lambda}\sum_{i: \sigma_i = 0}  \ketbra{i} + \sum_{n=0}^\infty(-\lambda)^n\sum_{i: \sigma_i > 0} \frac{1}{\sigma_i^{n+1}} \ketbra{i}\\
&= \frac{1}{\lambda}\qty[\bI - \bA\bA^+] + \sum_{n=0}^\infty (-\lambda)^n(\bA^+)^{n+1}.
\end{aligned}
\end{equation}

\subsection{Zero-Regularization Limit (\texorpdfstring{$\lambda \rightarrow 0$}{Lambda=0})}

Finally, we find approximate forms for the susceptibilities in Eq.~\eqref{eq:exact} by taking the limit in which the regularization constant goes to zero, $\lambda \rightarrow 0$.
Using the above identity, the susceptibilities then become
\begin{equation}
\mqty(\pdv{\vbu}{\vbf} & \pdv{\vbu}{\vbt}\\
\pdv{\vbell}{\vbf} & \pdv{\vbell}{\vbt}) = 
\mqty(  \bH^+ + \frac{1}{\lambda}\bP^{\mathrm{ZEM}}_H &  \qty[ \bH^+ + \frac{1}{\lambda}\bP^{\mathrm{ZEM}}_H] \bC^T\\
 \bC\qty[ \bH^+ + \frac{1}{\lambda}\bP^{\mathrm{ZEM}}_H] &  \bC\qty[ \bH^+ + \frac{1}{\lambda}\bP^{\mathrm{ZEM}}_H]\bC^T).
\end{equation}
where $\bP^{\mathrm{ZEM}}_H = \bI - \bH\bH^+$ is a projector onto the zero-energy modes in the kernel of $\bH$.

\setcounter{theorem}{0}
\setcounter{corollary}{0}

\section{Theorems and Proofs}\label{app:proofs}

This section contains proofs for the theorems presented in the main text. For convenience, we first define the following vector spaces:
\begin{enumerate}[(i)]
\item Constraint (component/bond) space: $\mathbb{R}^{N_\mathrm{c}}$
\item Degree-of-freedom (node) space: $\mathbb{R}^{N_\mathrm{dof}}$
\item Space of states of self-stress (SSSs):
\begin{equation*}
\mathcal{V}^{\mathrm{SSS}} \equiv \ker\qty(\bC^T) = \qty{\vbt\in\mathbb{R}^{N_\mathrm{c}} \middle| \bC^T\vbt = 0 }
\end{equation*}
\item Space of linear zero modes (LZMs): 
\begin{equation*}
\mathcal{V}^{\mathrm{LZM}} \equiv \ker\qty(\bC) = \qty{\vbu\in\mathbb{R}^{N_\mathrm{dof}} \middle| \bC\vbu = 0 }
\end{equation*}
\item Space of zero-displacement tensions (ZDTs): 
\begin{equation*}
\mathcal{V}^{\mathrm{ZDT}} \equiv \ker\qty(\pdv{\vbu}{\vbt}) =  \qty{\vbt\in\mathbb{R}^{N_\mathrm{c}} \middle| \pdv{\vbu}{\vbt}\vbt = 0 }
\end{equation*}
\item Space of linear zero-extension forces (LZEFs): 
\begin{equation*}
\mathcal{V}^{\mathrm{LZEF}} \equiv \ker\qty(\pdv{\vbell}{\vbf}) = \qty{\vbf\in\mathbb{R}^{N_\mathrm{dof}}\middle| \pdv{\vbell}{\vbf}\vbf = 0 }
\end{equation*}
\end{enumerate}

\subsection{Proofs of Theorem~\ref{thm:isomorphic} and Corollary~\ref{cor:eigenmode}}

\begin{theorem}\label{thm:SIisomorphic}
Consider a discrete elastic system described by the Hamiltonian in Eq.~\eqref{eq:hamiltonian}. The following statements are true:
\begin{enumerate}[(i)]
\item The ZDTs and SSSs are equivalent.
\item The LZEFs are isomorphic to the LZMs, with each LZEF $\vbfLZEF$ coupling to a unique LZM $\vbuLZM$ according to the bijective map
\begin{align}
\vbuLZM = \pdv{\vbu}{\vbf}\vbfLZEF.\label{eq:SIisomorphism}
\end{align}
\end{enumerate}
\end{theorem}

\begin{proof}
To start, we prove statement (i). Consider a SSS $\vbtSSS$ that by definition, lies in the kernel of $\bC^T$,
\begin{equation}
0 = \bC^T\vbtSSS.
\end{equation}
Multiplying both sides of this equation by ${\partial \vbu / \partial\vbf}$, we obtain
\begin{equation}
\begin{aligned}
0 &= \pdv{\vbu}{\vbf}\bC^T\vbtSSS\\
&= \pdv{\vbu}{\vbt}\vbtSSS,
\end{aligned}
\end{equation}
where in the second equality we have used the expression for ${\partial \vbu / \partial\vbt}$ in Eq.~\eqref{eq:suscept}.
Therefore, $\vbtSSS$ lies in the kernel of ${\partial \vbu / \partial\vbt}$ and is a ZDT.
Since this holds for every SSS, it must be that $\mathcal{V}^{\mathrm{SSS}}\subseteq \mathcal{V}^{\mathrm{ZDT}}$.

To prove the opposite direction, consider a ZDT $\vbtZDT$ that by definition, lies in the kernel of ${\partial \vbu / \partial\vbt}$,
\begin{equation}
0 = \pdv{\vbu}{\vbt}\vbtZDT.
\end{equation}
Because ${\partial \vbu / \partial\vbf}$ is full rank, it is invertible. 
Multiplying both sides of this equation by $[\partial \vbu / \partial\vbf]^{-1}$, we obtain
\begin{equation}
\begin{aligned}
0 &=  \qty[\pdv{\vbu}{\vbf}]^{-1} \pdv{\vbu}{\vbt}\vbtZDT\\
&= \qty[\pdv{\vbu}{\vbf}]^{-1} \pdv{\vbu}{\vbf}\bC^T\vbtZDT\\
&= \bC^T\vbtZDT,
\end{aligned}
\end{equation}
where in the second equality we have again used the expression for ${\partial \vbu / \partial\vbt}$ in Eq.~\eqref{eq:suscept}.
Therefore $\vbtZDT$ lies in the kernel of $\bC^T$ and is an SSS.
Since this holds for every ZDT, it must be that $\mathcal{V}_{ZDT}\subseteq \mathcal{V}_{\mathrm{SS}}$.
Combining this with the previous result, we conclude that the two spaces are equal, $\mathcal{V}^{\mathrm{ZDT}} = \mathcal{V}^{\mathrm{SSS}}$.

Next, we prove statement (ii). Consider a LZEF $\vbfLZEF$ which by definition, lies in the kernel of ${\partial\vbell / \partial\vbf}$,
\begin{equation}
0 = \pdv{\vbell}{\vbf}\vbfLZEF.
\end{equation}
Because the susceptibility ${\partial \vbu / \partial\vbf}$ is full rank, we may map $\vbfLZEF$ to the unique nonzero displacement that it creates when applied to the system,
\begin{equation}
\vbu = \pdv{\vbu}{\vbf}\vbfLZEF.
\end{equation}
Multiplying both sides by $\bC$, we find
\begin{equation}
\begin{aligned}
\bC\vbu &= \bC\pdv{\vbu}{\vbf}\vbfLZEF\\
&= \pdv{\vbell}{\vbf}\vbfLZEF\\
&= 0,
\end{aligned}
\end{equation}
where in the second equality we have used the expression for ${\partial\vbell / \partial\vbf}$ from Eq.~\eqref{eq:suscept}.
We find that $\vbu$ lies in the kernel of $\bC$ and is, therefore, a LZM.
Therefore, every LZEF $\vbfLZEF$ can be mapped to the LZM $\vbuLZM$ it gives rise to via the relation
\begin{equation}
\vbuLZM = \pdv{\vbu}{\vbf}\vbfLZEF.
\end{equation}

To prove the opposite direction, consider a LZM $\vbuLZM$ which by definition, lies in the kernel of $\bC$,
\begin{equation}
0 = \bC \vbuLZM.
\end{equation}
Using  $[\partial \vbu / \partial\vbf]^{-1}$, we may map $\vbuLZM$ to the unique external force that creates it when applied to the system,
\begin{equation}
\vbf = \qty[\pdv{\vbu}{\vbf}]^{-1}\vbuLZM.
\end{equation} 
Multiplying both sides by ${\partial\vbell / \partial\vbf}$, we find
\begin{equation}
\begin{aligned}
\pdv{\vbell}{\vbf}\vbf &= \pdv{\vbell}{\vbf}\qty[\pdv{\vbu}{\vbf}]^{-1}\vbuLZM\\
&= \bC\pdv{\vbu}{\vbf}\qty[\pdv{\vbu}{\vbf}]^{-1}\vbuLZM\\
&= \bC \vbuLZM\\
&=0,
\end{aligned}
\end{equation}
where in the second equality we again have used the expression for ${\partial\vbell / \partial\vbf}$ from Eq.~\eqref{eq:suscept}.
We find that $\vbf$ lies in the kernel of ${\partial\vbell / \partial\vbf}$ and is therefore a LZEF.
Therefore, every  LZM $\vbuLZM$ can be mapped to the LZEF $\vbfLZEF$ that gives rise to it via the relation
\begin{equation}
\vbfLZEF = \qty[\pdv{\vbu}{\vbf}]^{-1}\vbuLZM.
\end{equation}
Combing this with the previous result, we conclude that the two spaces are isomorphic,  $\mathcal{V}^{\mathrm{LZEF}} \cong \mathcal{V}^{\mathrm{SSS}}$, related by the bijective map
\begin{align}
\pdv{\vbu}{\vbf}:&  \mathcal{V}^{\mathrm{LZEF}}\rightarrow\mathcal{V}^{\mathrm{SSS}}.
\end{align}
\end{proof}

\begin{corollary}\label{cor:SIeigenmode}
Consider a LZM $\vbuLZM$ and a LZEF $\vbfLZEF$ related by Eq.~\eqref{eq:SIisomorphism}. The two modes are parallel, $\vbuLZM \parallel \vbfLZEF$, if and only if they are eigenmodes of Hessian.
\end{corollary}

\begin{proof}
This is a straightforward consequence of Eq.~\eqref{eq:SIisomorphism}. If the two modes are parallel, then they must correspond to an eigenmode of the susceptibility ${\partial \vbu / \partial \vbf}$. Since the susceptibility is just the inverse of the Hessian plus a term proportional to the identity, it will have the same eigenmodes as the Hessian. Conversely, if the two modes correspond to an eigenmode of the Hessian, then they will be parallel.
\end{proof}

\subsection{Linear Invariance Relations}
In this section we derive the matrix invariance relations in Eqs.~\eqref{eq:invarC}-\eqref{eq:invardldf}, along with invariance relations for many other matrix operators that appear in the main text.
We formulate these derivations into Lemmas that we will later use to prove Theorem~\ref{thm:symmetry} and Corollary~\ref{cor:subspace}.
Throughout these proofs, we use repeated indices to imply sums and will often suppress the subscript for matrix representations that indicate their group. We will also use the fact that the matrix representations are unitary and real (because they simply permute the bonds or nodes and rotate vectors), allowing us to write
\begin{equation}
\begin{gathered} 
\bD(g)^T = \bD(g)^\dag = \bD(g)^{-1} = \bD(g^{-1})
\end{gathered}
\end{equation} 
for any of our representations.

\begin{lemma}\label{lemma:geominvar}
If a system displays geometric equivariance according to Eq.~\eqref{eq:invarell}, then the following invariance relations hold for the compatibility matrix and its derivative:
\begin{align}
\bC &= \bDell_G(g^{-1})\bC\bDu_G(g)\\
\pdv{\ell_\alpha}{u_i}{u_j}  &=  \Dell_{G, \alpha\beta}(g^{-1})\pdv{\ell_\beta}{u_k}{u_l}\Du_{G, ki}(g)\Du_{G, lj}(g).
\end{align}
\end{lemma}

\begin{proof}
First, we take the derivative of Eq.~\eqref{eq:invarell} with respect to $\vbu$. 
Defining $\vbu' = \bDu(g)\vbu$, we find
\begin{equation}
\begin{aligned}
\pdv{\ell_\alpha(\vbu)}{u_i} &= \Dell_{\alpha\beta}(g^{-1})\pdv{\ell_\beta(\vbu')}{u_i} \\
&= \Dell_{\alpha\beta}(g^{-1})\pdv{\ell_\beta(\vbu')}{u'_j}\pdv{u'_j}{u_i}\\
&= \Dell_{\alpha\beta}(g^{-1})\pdv{\ell_\beta(\vbu')}{u'_j}\Du_{ji}(g).\label{eq:SIinvardelldu}
\end{aligned}
\end{equation}
Evaluating at $\vbu=\vbu'=0$, we find that $\bC$ obeys the invariance relation
\begin{equation}
\bC = \bDell(g^{-1})\bC\bDu(g).
\end{equation}

Taking an additional derivative of Eq.~\eqref{eq:SIinvardelldu} with respect to $\vbu$, we find an invariance relation for the second derivative of $\ell_\alpha$,
\begin{equation}
\begin{aligned}
\pdv{\ell_\alpha(\vbu)}{u_i}{u_j} &= \Dell_{\alpha\beta}(g^{-1})\pdv{\ell_\beta(\vbu')}{u'_k}{u_j}\Du_{ki}(g)\\
&= \Dell_{\alpha\beta}(g^{-1})\pdv{\ell_\beta(\vbu')}{u'_k}{u'_l}\Du_{ki}(g)\pdv{u'_l}{u_j}\\
&= \Dell_{\alpha\beta}(g^{-1})\pdv{\ell_\beta(\vbu')}{u'_k}{u'_l}\Du_{ki}(g)\Du_{lj}(g).
\end{aligned}
\end{equation}
Evaluating at $\vbu=\vbu' = 0$, we find
\begin{equation}
\pdv{\ell_\alpha}{u_i}{u_j}  =  \Dell_{\alpha\beta}(g^{-1})\pdv{\ell_\beta}{u_k}{u_l}\Du_{ki}(g)\Du_{lj}(g).
\end{equation}
\end{proof}

\begin{lemma}\label{lemma:energyinvar}
If a system displays energetic invariance according to Eq.~\eqref{eq:invarE}, then the following invariance relations hold for the local energetics:
\begin{align}
\vbtp &=  \bDell_H(h)\vbtp\\
\bK &= \bDell_H(h^{-1})\bK \bDell_H(h).
\end{align}
\end{lemma}

\begin{proof}
First, we take a derivative of Eq.~\eqref{eq:invarE} with respect to $\vbell$.
Defining $\delta \vbell' = \bDell(h)\delta \vbell$, we find 
\begin{equation}
\begin{aligned}
\pdv{E(\vbell)}{\ell_\alpha} &= \pdv{E(\vbell_0 + \delta\vbell')}{\ell_\alpha}\\
&=  \pdv{E(\vbell_0 + \delta\vbell')}{\ell'_\beta}\pdv{\ell'_\beta}{\ell_\alpha}\\
&= \pdv{E(\vbell_0 + \delta\vbell')}{\ell'_\beta} \Dell_{\beta\alpha}(h).\label{eq:SIinvardEdell}
\end{aligned}
\end{equation}
Evaluating at $\delta \vbell=\delta \vbell' = 0$, this gives us an invariance relation for the prestress $\vbtp$,
\begin{equation}
\vbtp =  \bDell(h)\vbtp
\end{equation}
(since this is true for all $h$, we replaced $h^{-1}$ with $h$).

Taking an additional derivative of Eq.~\eqref{eq:SIinvardEdell} with respect to $\vbell$, we find
\begin{equation}
\begin{aligned}
\pdv{E(\vbell)}{\ell_\alpha}{\ell_\beta} &=  \pdv{E(\vbell_0 + \delta\vbell')}{\ell'_\gamma}{\ell_\beta} \Dell_{\gamma\alpha}(h)\\
&=  \pdv{E(\vbell_0 + \delta\vbell')}{\ell'_\gamma}{\ell'_\delta} \Dell_{\gamma\alpha}(h)\Dell_{\beta\delta}(h).
\end{aligned}
\end{equation}
At $\delta \vbell=\delta \vbell' = 0$, this also gives us an invariance relation for the stiffness matrix $\bK$,
\begin{equation}
\bK = \bDell(h^{-1})\bK \bDell(h).
\end{equation}
\end{proof}

\begin{lemma}\label{lemma:bothinvar}
If a system displays obeys both the  geometric and energetic symmetry conditions, Eq.~\eqref{eq:invarell} and \eqref{eq:invarE}, then the Hessian transforms as
\begin{align}
\bH &= \bDu_{G\cap H}(g^{-1})\bH \bDu_{G\cap H}(g)\label{eq:SIinvarH}
\end{align}
and the susceptibilities transform as
\begin{align}
\pdv{\vbu}{\vbf} &= \bDu_{G\cap H}(g^{-1})\pdv{\vbu}{\vbf} \bDu_{G\cap H}(g),\\
\pdv{\vbu}{\vbt} &= \bDu_{G\cap H}(g^{-1})\pdv{\vbu}{\vbt} \bDell_{G\cap H}(g),\\
\pdv{\vbell}{\vbf} &= \bDell_{G\cap H}(g^{-1})\pdv{\vbell}{\vbf} \bDu_{G\cap H}(g),\label{eq:SIinvardldf}\\
\pdv{\vbell}{\vbt} &= \bDell_{G\cap H}(g^{-1})\pdv{\vbell}{\vbt} \bDell_{G\cap H}(g).
\end{align}
\end{lemma}
\begin{proof}
There are two alternative routes to derive Eq.~\eqref{eq:SIinvarH}.
The first is to combine the invariance relations from Lemmas~\ref{lemma:geominvar} and \ref{lemma:energyinvar} and use the formula for the Hessian in Eq.~\eqref{eq:Hessian}.
Alternatively, we may take derivatives of the energy directly via Eq.~\eqref{eq:invarE} with respect to $\vbu$. Following this route, we first combine Eqs~\eqref{eq:invarell} and \eqref{eq:invarE} to find an invariance relation for the energy under transformation of the displacements. We also restrict ourselves to representation matrices in the intersection $G\cap H$. We find
\begin{equation}
\begin{aligned}
E(\vbell(\vbu)) &= E\qty(\vbell_0 + \bDell(g)\delta \vbell\qty(\vbu))\\
&= E\qty(\vbell_0 + \delta \vbell\qty(\bDu(g)\vbu))\\
&= E\qty(\vbell\qty(\bDu(g)\vbu)).
\end{aligned}
\end{equation}

Defining $\vbu' = \bDu(g)\vbu$, we take two derivatives to find
\begin{equation}
\begin{aligned}
\pdv{E(\vbell(\vbu))}{u_i}{u_j} &=  \pdv{E\qty(\vbell\qty(\vbu'))}{u_i}{u_j}\\
&= \pdv{E\qty(\vbell\qty(\vbu'))}{u_k'}{u_l'}\pdv{u_k'}{u_i}\pdv{u_l'}{u_j}\\
&= \pdv{E\qty(\vbell\qty(\vbu'))}{u_k'}{u_l'}\Du_{ki}(g)\Du_{lj}(g).
\end{aligned}
\end{equation}
Evaluating at $\vbu=\vbu'=0$, we get
\begin{equation}
\bH = \bDu(g^{-1})\bH \bDu(g).
\end{equation}

Next, we derive the invariance relations for the susceptibilities by combing the invariance relations derived up to this point with the exact formulas for the susceptibilities in Eq.\eqref{eq:exact}.
First, we compute the invariance relation for the node-space susceptibility ${\partial \vbu / \partial\vbf}$,
\begin{equation}
\begin{aligned}
\pdv{\vbu}{\vbf} &= \qty[\bH + \lambda \bI]^{-1}\\
&= \qty[\bDu(g^{-1})\bH\bDu(g) + \lambda \bI]^{-1}\\
&= \qty[\bDu(g^{-1})\qty(\bH+ \lambda \bI)\bDu(g)]^{-1}\\
&= \bDu(g)^{-1} \qty[\bH + \lambda \bI]^{-1}\bDu(g^{-1})^{-1}\\
&= \bDu(g^{-1})\qty[\bH + \lambda \bI]^{-1}\bDu(g)\\
&= \bDu(g^{-1})\pdv{\vbu}{\vbf}\bDu(g).
\end{aligned}
\end{equation}

To compute the invariance relation for ${\partial \vbu / \partial\vbt}$, we need an invariance relation for $\bC^T$,
\begin{equation}
\begin{aligned}
\bC^T &= \qty(\bDell(g^{-1})\bC\bDu(g))^T\\
 &= \bDu(g)^T\bC^T \bDell(g^{-1})^T\\
&= \bDu(g^{-1})\bC^T \bDell(g).
\end{aligned}
\end{equation}
Combining this with the invariance relation for ${\partial \vbu / \partial\vbf}$, we find
\begin{equation}
\begin{aligned}
\pdv{\vbu}{\vbt} &= \pdv{\vbu}{\vbf}\bC^T\\
&= \bDu(g^{-1})\pdv{\vbu}{\vbf}\bDu(g)\bDu(g^{-1})\bC^T \bDell(g)\\
&=  \bDu(g^{-1})\pdv{\vbu}{\vbf}\bC^T \bDell(g)\\
&= \bDu(g^{-1})\pdv{\vbu}{\vbt} \bDell(g).
\end{aligned}
\end{equation}

Similarly, we derive the invariance relation for ${\partial\vbell / \partial\vbf}$,
\begin{equation}
\begin{aligned}
\pdv{\vbell}{\vbf} &= \bC\pdv{\vbu}{\vbf}\\
&= \bDell(g^{-1})\bC\bDu(g)\bDu(g^{-1})\pdv{\vbu}{\vbf}\bDu(g)\\
&= \bDell(g^{-1})\bC\pdv{\vbu}{\vbf}\bDu(g)\\
&= \bDell(g^{-1}) \pdv{\vbell}{\vbf}\bDu(g).
\end{aligned}
\end{equation}

Finally, we derive the invariance relation for ${\partial \vbell / \partial \vbt}$,
\begin{equation}
\begin{aligned}
\pdv{\vbell}{\vbt} &= \bC\pdv{\vbu}{\vbt}\\
&= \bDell(g^{-1})\bC\bDu(g) \bDu(g^{-1})\pdv{\vbu}{\vbf}\bDu(g)\\
&= \bDell(g^{-1})\bC\pdv{\vbu}{\vbt}\bDell(g)\\
&= \bDell(g^{-1}) \pdv{\vbell}{\vbt}\bDell(g).
\end{aligned}
\end{equation}
\end{proof}

\subsection*{Proofs of Theorem~\ref{thm:symmetry} and Corollary~\ref{cor:subspace}}

\begin{theorem}\label{thm:SIsymmetry}
Consider a discrete elastic system described by the Hamiltonian in Eq.~\eqref{eq:hamiltonian}. 
Suppose the geometry exhibits the symmetry of a group $G$ [Eq.~\eqref{eq:invarell}], the energy exhibits the symmetry of a group $H$ [Eq.~\eqref{eq:invarE}], and the elastic response obeys the symmetry of the group $G\cap H$, all with the corresponding matrix representations defined above. Then the following is true:
\begin{enumerate}[(a)]
\item The LZMs obey the symmetry of the geometric symmetry group $G$, forming an invariant subspace under the action of $\bDu_G(g)$ such that for every LZM $\vbuLZM$ and $g\in G$, $\bDu_G(g)\vbuLZM$ is also a LZM.
\item The LZEFs obey the symmetry of elastic response symmetry group $G\cap H$, forming an invariant subspace under the action of $\bDu_{G\cap H}(g)$ such that for every LZEF $\vbfLZEF$ and $g\in G \cap H$, $\bDu_{G\cap H}(g)\vbfLZEF$ is also a LZEF.
\end{enumerate}
\end{theorem}

\begin{proof}
To prove (a), suppose there exists a LZM $\vbu$ which lies in the kernel of $\bC$,
\begin{equation}
\bC\vbu=0.
\end{equation}
Now consider the mode $\bDu_G(g)\vbu$ for an arbitrary element $g\in G$.
Acting on this mode with $\bC$ and applying Lemma~\ref{lemma:geominvar},  we find
\begin{equation}
\bC\bDu_G(g)\vbu = \bDell_G(g)\bC \vbu = 0.
\end{equation}
We conclude that $\bDu_G(g)\vbu$ is also a LZM.
Therefore, the LZMs form an invariant subspace of $\mathbb{R}^{N_{\mathrm{dof}}}$ under $G$.

Next, we prove (b). Suppose there exists a LZEF $\vbf$ which lies in the kernel of ${\partial\vbell / \partial\vbf}$,
\begin{equation}
\pdv{\vbell}{\vbf} \vbf = 0.
\end{equation}
Now consider the mode $\bDu_{G\cap H}(g)\vbf$ for an arbitrary element $g\in G\cap H$.
Acting on this mode with ${\partial\vbell / \partial\vbf}$ and applying Lemma~\ref{lemma:bothinvar}, we find
\begin{equation}
\pdv{\vbell}{\vbf} \bDu_{G\cap H}(g)\vbf = \bDell_{G\cap H}(g)\pdv{\vbell}{\vbf}  \vbf = 0.
\end{equation}
We conclude that $\bDu_{G\cap H}(g)\vbf$ is also a LZEF.
Therefore, the LZEFs form an invariant subspace of $\mathbb{R}^{N_{\mathrm{dof}}}$ under $G\cap H$. 
\end{proof}

\begin{corollary}\label{cor:SIsubspace}
Consider a discrete elastic system that satisfies Theorem~\ref{thm:SIsymmetry} with a set of LZMs and LZEFs related by Eq.~\eqref{eq:isomorphism},
spanning the vector spaces $\mathcal{V}^{\mathrm{LZM}}$ and $\mathcal{V}^{\mathrm{LZEF}}$, respectively.
Define the subspace $\mathcal{V}^\Gamma$ spanned by all vectors that transform according to some IRREP $\Gamma$ under the action of the elastic response node space representation $\bDu_{G\cap H}$ [satisfying Eq.~\eqref{eq:Dinvarspaces}],
\begin{equation}
\mathcal{V}^\Gamma = \qty{ \sum_{s=1}^{n_\Gamma}\sum_{i=1}^{d_\Gamma} a_{\Gamma_s, i}\ket{\Gamma_s, i} \middle| a_{\Gamma_s, i} \in \mathbb{R}},
\end{equation}
where $\ket{\Gamma_s, i}$ is an arbitrary complete orthonormal basis that block-diagonalizes $\bDu_{G\cap H}$ [Eq.~\eqref{eq:Dblockdiag}].

If either the LZM subspace $\mathcal{V}^{\mathrm{LZM}}$ or LZEF subspace $\mathcal{V}^{\mathrm{LZEF}}$ is equal to a subspace $\mathcal{V}^\Gamma$ for some IRREP $\Gamma$, then the LZM sand LZEF subspaces are equal, ${\mathcal{V}^{\mathrm{LZM}} = \mathcal{V}^{\mathrm{LZEF}}}$.
\end{corollary}

\begin{proof}
Let us choose the basis $\ket{\Gamma_s, i}$ to be the eigenmodes of the Hessian $\bH$.
Now if the LZM subspace $\mathcal{V}^{\mathrm{LZM}}$ is equal to $\mathcal{V}^\Gamma$ some IRREP $\Gamma$, then we may write the LZMs as eigenmodes of the Hessian. The same is true for the LZEF subspace $\mathcal{V}^{\mathrm{LZEF}}$. In either case, we may appeal to Corollary~\ref{cor:SIeigenmode} to conclude that the LZMs and LZEFs span the same space as eigenmodes of the Hessian.
\end{proof}

\section{Irreducible Representations and Mode Structure}\label{app:irreps}

In this section, we provide a partial derivation of Eqs.~\eqref{eq:Dinvarspaces}-\eqref{eq:Hblockdiag}.
In doing so, we illustrate how knowledge of the irreducible representations of a symmetry group can provide information about the mode structure of operators that transform under that group. 
For a more thorough treatment, we recommend Ref.~\onlinecite{Arovas2022}.

To begin,  we define the irreducible representations, or IRREPs, of $G$, indexed by $\Gamma$, as ${\bD^\Gamma: G\rightarrow \mathrm{GL}(\mathcal{V}^\Gamma)}$ where $\mathcal{V}^\Gamma$ is a vector space of dimension $d_\Gamma$.
As a reminder, an IRREP is a representation under which $\mathcal{V}$ contains no nontrivial invariant subspaces.
A subspace $\mathcal{V}'\subset \mathcal{V}$ is invariant if 
${\bD(g)\ket{a} \in \mathcal{V}'}$ for all $\ket{a} \in  \mathcal{V}'$ and for all $g\in G$.
A nontrivial invariant subspace is one other than $\mathcal{V}$ or the null vector $\{0\}$.

Next, let ${\bD:G\rightarrow  \mathrm{GL}(\mathcal{V})}$ be a representation of $G$ that acts on a vector space $\mathcal{V}$ of dimension $N$.
Using the IRREPs, we may construct the set of operators
\begin{equation}
\begin{gathered}
\bP^{\Gamma, ij} = \frac{d_\Gamma}{N_G}\sum_{g\in G}D^\Gamma_{ji}(g^{-1})\bD(g)
\end{gathered}
\end{equation}
where $\Gamma$ ranges through the different IRREPs and $i$ and $j$ range from 1 to $d_\Gamma$.
These operators can be used to create projectors onto the invariant subspaces of $\mathcal{V}$ that transform according to each IRREP under the action of $\bD$.

To show this, we will first prove that these operators are orthogonal to one another and are complete.
To prove orthogonality, we will need the following identity:
\begin{equation}
\begin{aligned}
\bP^{\Gamma, ij}\bD(g) &= \frac{d_\Gamma}{N_G}\sum_{g'\in G} D^\Gamma_{ji}(g'^{-1})\bD(g')\bD(g)\\
&= \frac{d_\Gamma}{N_G}\sum_{g'\in G} D^\Gamma_{ji}(g'^{-1})\bD(g'g)\\
&= \frac{d_\Gamma}{N_G}\sum_{g''\in G} D^\Gamma_{ji}(gg''^{-1})\bD(g'')\\
&= \sum_{k=1}^{d_\Gamma}D^\Gamma_{jk}(g)\frac{d_\Gamma}{N_G}\sum_{g''\in G} D^\Gamma_{ki}(g''^{-1})\bD(g'')\\
&= \sum_{k=1}^{d_\Gamma} D^\Gamma_{jk}(g) \bP^{\Gamma, ik}.
\end{aligned}
\end{equation}
We will also need the Great Orthogonality Theorem for IRREPs~\cite{Arovas2022},
\begin{equation}
\frac{1}{N_G}\sum_{g\in G}D^\Gamma_{ik}(g^{-1})D^{\Gamma'}_{lj}(g) = \frac{1}{d_\Gamma}\delta_{\Gamma\Gamma'}\delta_{ij}\delta_{kl}.
\end{equation}
Using these identities, we may now write
\begin{equation}
\begin{aligned}
\bP^{\Gamma, ij}\bP^{\Gamma', kl} &= \frac{d_{\Gamma'}}{N_G}\sum_{g'\in G} D^{\Gamma'}_{lk}(g'^{-1})\bP^{\Gamma, ij}\bD(g')\\
&= \sum_{m=1}^{d_\Gamma} \bP^{\Gamma, im} \frac{d_{\Gamma'}}{N_G}\sum_{g'} D^{\Gamma'}_{lk}(g'^{-1}) D^\Gamma_{jm}(g')\\
&=\sum_{m=1}^{d_\Gamma} \bP^{\Gamma, im} \delta_{\Gamma\Gamma'}\delta_{lm}\delta_{jk} \\
&= \delta_{\Gamma\Gamma'} \delta_{jk}\bP^{\Gamma', il}.
\end{aligned}
\end{equation}
We see that these operators are indeed orthogonal to one another.
Furthermore, the operators $\bP^{\Gamma, ii}$ are orthogonal projectors.

To prove completeness, we will need an identity for group character orthogonality~\cite{Arovas2022}. We define the group characters as $\chi(g) = \tr \bD(g)$. In addition, we use the notation $\mathcal{C}(g)$ to denote the conjugacy class of $g$. All elements in $\mathcal{C}(g)$ have the same character, so we will also use $\chi(\mathcal{C}(g))$ to indicate character of elements in the conjugacy class of $g$. Now we have the identity,
\begin{equation}
\sum_\Gamma \chi^\Gamma(\mathcal{C}(g^{-1}))\chi^\Gamma(\mathcal{C}(g')) = \frac{N_G}{N_{\mathcal{C}(g)}}\delta_{\mathcal{C}(g)\mathcal{C}(g')}.
\end{equation}
We use this to prove completeness,
\begin{equation}
\begin{aligned}
\sum_\Gamma \sum_{j=1}^{d_\Gamma} \bP^{\Gamma, jj} &= \sum_\Gamma\sum_{j=1}^{d_\Gamma} \frac{d_\Gamma}{N_G}\sum_{g\in G}D^\Gamma_{jj}(g^{-1})\bD(g)\\
&= \sum_{g\in G}\bD(g)  \frac{1}{N_G} \sum_\Gamma  \chi^\Gamma(g^{-1}) d_\Gamma\\
&= \sum_{g\in G}\bD(g) \frac{1}{N_G} \sum_\Gamma \chi^\Gamma(g^{-1}) \chi^\Gamma(I)\\
&= \sum_{g\in G}\bD(g) \frac{1}{N_G} \sum_\Gamma \chi^\Gamma(\mathcal{C}(g^{-1})) \chi^\Gamma(\mathcal{C}(I))\\
&=  \sum_{g\in G}\bD(g) \frac{1}{N_{\mathcal{C}(I)}}\delta_{\mathcal{C}(g)\mathcal{C}(I)}  \\
&=  \sum_{g\in G}\bD(g) \delta_{gI}  \\
&= \bD(I)\\
&= \bI.
\end{aligned}
\end{equation}
In the third line, we identified $d_\Gamma$ as the character of the identity element for each IRREP.

Now that we have defined these operators, we use them to construct a basis for the vector space $\mathcal{V}$. 
We construct the orthonormal basis $\ket{\Gamma_s, i}$ such that the operators decompose as follows:
\begin{equation}
\bP^{\Gamma, ij} = \sum_{s=1}^{n_\Gamma} \ket{\Gamma_s, i}\bra{\Gamma_s, j},
\end{equation}
such that 
\begin{equation}
\begin{gathered}
\braket{\Gamma_s, i}{\Gamma_{s'}, j} = \delta_{\Gamma\Gamma'}\delta_{ss'}\delta_{ij},\\
\sum_\Gamma\sum_{s=1}^{n_\Gamma}\sum_{j=1}^{d_\Gamma}\ketbra{\Gamma_s, j} = \bI.
\end{gathered}
\end{equation}
Each basis element $\ket{\Gamma_s, i}$ in index by both an IRREP $\Gamma$ and a mode number $i=1,\ldots,d_\Gamma$. Furthermore, we define an additional index $s=1,\ldots,n_\Gamma$ where $n_\Gamma$ is the rank of the operator $\bP^{\Gamma, ij}$.
It is straightforward to see that these operators obey their proper orthogonality and completeness properties when written in this way. 

We note that this decomposition is not unique. In particular, we may arbitrarily rotate between vectors with different values of $s$ to acquire equally valid sets of basis vectors, as long as we do not mix between different IRREPs $\Gamma$ and mode numbers $i$, e.g.,
\begin{equation}
\ket{\Gamma_s, i} \rightarrow \sum_{s'}S_{ss'}\ket{\Gamma_s, i}
\end{equation}
where $\bS$ is an $n_\Gamma\times n_\Gamma$ orthogonal matrix $\bS^T\bS=\bS\bS^T=\bI$.

Now we show that in this basis, the matrix representation $\bD$ acquires the block-diagonal form in Eq.~\eqref{eq:Dblockdiag},
\begin{equation}
\begin{aligned}
\bD(g) &= \sum_\Gamma\sum_{j=1}^{d_\Gamma} \bP^{\Gamma, jj}\bD(g)\\
&=  \sum_\Gamma\sum_{j,k=1}^{d_\Gamma} D^\Gamma_{jk}(g) \bP^{\Gamma, jk}\\
&= \sum_\Gamma \sum_{s=1}^{n_\Gamma} \sum_{j,k=1}^{d_\Gamma} \ket{\Gamma_s, j} D^\Gamma_{jk}(g)\bra{\Gamma_s, k}.
\end{aligned}
\end{equation}
We see that we can interpret $n_\Gamma$ as the number of times the IRREP $\bD^\Gamma$ appears within our representation $\bD$ and that $s$ labels each copy.
Furthermore, we see that in this basis, each collection of basis elements for fixed $\Gamma_s$ forms an invariant subspace under the action of $\bD$ [Eq.~\eqref{eq:Dinvarspaces}],
\begin{equation}
\begin{aligned}
\bD(g)\ket{\Gamma_s, i} &= \sum_{\Gamma'} \sum_{s=1}^{n_{\Gamma'}} \sum_{j,k=1}^{d_{\Gamma'}} \ket{{\Gamma'}_s, j} D^{\Gamma'}_{jk}(g)\bra{\Gamma'_s, k}\ket{\Gamma_s, i}\\
&= \sum_{j=1}^{d_\Gamma} \ket{\Gamma_s, j} D^\Gamma_{ji}(g)
\end{aligned}
\end{equation}

Now we use our basis to diagonalize the Hessian $\bH$. First, we act with $\bP^{\Gamma, jj}$ onto the Hessian to find
\begin{equation}
\begin{aligned}
\bP^{\Gamma, jj}\bH &= \frac{1}{N_g}\sum_{g\in G} \bP^{\Gamma, jj} \bD(g^{-1})\bH\bD(g)\\
&= \frac{1}{N_g}\sum_{g\in G} \sum_{k=1}^{d_\Gamma} D^\Gamma_{jk}(g^{-1}) \bP^{\Gamma, jk}\bH\bD(g)\\
&= \sum_{k=1}^{d_\Gamma} \bP^{\Gamma, jk}\bH \frac{1}{N_g}\sum_{g\in G} D^\Gamma_{jk}(g^{-1}) \bD(g)\\
&= \frac{1}{d_\Gamma}\sum_{k=1}^{d_\Gamma} \bP^{\Gamma, jk}\bH \bP^{\Gamma, kj}
\end{aligned}
\end{equation}
From this, we sum over $\Gamma$ and $j$ to find
\begin{equation}
\begin{aligned}
\bH &= \sum_\Gamma\sum_{j=1}^{d_\Gamma} \bP^{\Gamma, jj}\bH\\
&=  \frac{1}{d_\Gamma}\sum_\Gamma\sum_{j,k=1}^{d_\Gamma} \bP^{\Gamma, jk}\bH \bP^{\Gamma, kj}\\
&= \frac{1}{d_\Gamma}\sum_\Gamma \sum_{ss'}  \sum_{j,k=1}^{d_\Gamma} \ket{\Gamma_s, j}  \mel{\Gamma_s, k}{\bH}{\Gamma_{s'}, k}\bra{\Gamma_{s'}, j}\\
&= \sum_\Gamma \sum_{s,s'=1}^{n_\Gamma}\sum_{j=1}^{d_\Gamma} \ket{\Gamma_s, j} H^\Gamma_{ss'}\bra{\Gamma_{s'}, j}\label{eq:Halmostdiag}
\end{aligned}
\end{equation}
where we have defined the $n_\Gamma\times n_\Gamma$ matrix
\begin{equation}
H^\Gamma_{ss'} = \frac{1}{d_\Gamma} \sum_{k=1}^{d_\Gamma} \mel{\Gamma_s, k}{\bH}{\Gamma_{s'}, k}.
\end{equation}
To fully diagonalize $\bH$, we perform an eigendecomposition of $\bH^\Gamma$,
\begin{equation}
H^\Gamma_{ss'} = \sum_{t=1}^{n_\Gamma} S_{st}\omega^2_{\Gamma_t}S^T_{ts'}
\end{equation}
where $\omega^2_{\Gamma_t}$ are the eigenvalues of $\bH^\Gamma$ and the columns of $\bS$ contain the eigenvectors, which may be chosen to be orthonormal because $\bH^\Gamma$ is symmetric.
This means that $\bS$ is an orthogonal matrix, so we may use it to define a new set of orthonormal basis vectors, 
\begin{equation}
\overline{\ket{\Gamma_s, i}} \equiv \sum_{t=1}^{n_\Gamma} S^T_{st} \ket{\Gamma_t, i}.
\end{equation}
Substituting this into Eq.~\eqref{eq:Halmostdiag}, we see that $\bH$ is now fully diagonalized,
\begin{equation}
\begin{aligned}
\bH &= \sum_\Gamma \sum_{s,s'=1}^{n_\Gamma}\sum_{j=1}^{d_\Gamma} \ket{\Gamma_s, j} \qty(\sum_{t=1}^{n_\Gamma} S_{st}\omega^2_{\Gamma_t}S^T_{ts'})\bra{\Gamma_{s'}, j}\\
&= \sum_\Gamma \sum_{t=1}^{n_\Gamma}  \omega^2_{\Gamma_t} \sum_{j=1}^{d_\Gamma}\qty(\sum_{s=1}^{n_\Gamma} S^T_{ts}\ket{\Gamma_s, j}) \qty(\sum_{s'=1}^{n_\Gamma} \bra{\Gamma_{s'}, j}S_{s't})\\
&= \sum_\Gamma \sum_{t=1}^{n_\Gamma}  \omega^2_{\Gamma_t}\sum_{j=1}^{d_\Gamma}  \overline{\ketbra{\Gamma_t, j}}.
\end{aligned}
\end{equation}
The set of vectors $\overline{\ket{\Gamma_t, j}}$ for fixed $\Gamma$ spans the same vector space as the vectors $\ket{\Gamma_t, j}$. 
We can easily show that these vectors properly transform according to their respective IRREPs under the action of $\bD$,
\begin{equation}
\begin{aligned}
\bD(g)\overline{\ket{\Gamma_s, i}} &=\sum_{t=1}^{n_\Gamma}  \bD(g) S^T_{st}\ket{\Gamma_t, i}\\
&= \sum_{t=1}^{n_\Gamma}  S^T_{st} \sum_{j=1}^{d_\Gamma} \ket{\Gamma_t, j}D^\Gamma_{ji}(g)\\
&= \sum_{j=1}^{d_\Gamma} \overline{\ket{\Gamma_s, j}}D^\Gamma_{ji}(g).
\end{aligned}
\end{equation}

\section{Group Definitions, Irreducible Representations, Node Space Representations}

In this section, we define the groups used in the main text and report their irreducible representations. We also explain how we construct the node space representations for these groups used to classify the modes of the Hessian.

\subsection{Diamond Network}\label{app:diamond}

Here we describe the groups used to capture the symmetries of the diamond network in Figs.~\ref{fig:symbreak} and \ref{fig:levels}.

\subsubsection{\texorpdfstring{$D_4$}{D4} Group}

The geometric symmetry of diamond network in Figs.~\ref{fig:symbreak} and \ref{fig:levels} is described by the four-fold dihedral group $D_4$, with group presentation
\begin{equation}
D_4 = \langle r, \sigma \mid r^4=\sigma^2=(\sigma r)^2=1\rangle.
\end{equation}
where the generator $r$ corresponds to a $90^\circ$ counterclockwise rotation of the network about the center node, and $\sigma$ corresponds to reflection across the central vertical axis.

\begin{figure}[h]
\centering
\includegraphics[width=0.67\linewidth]{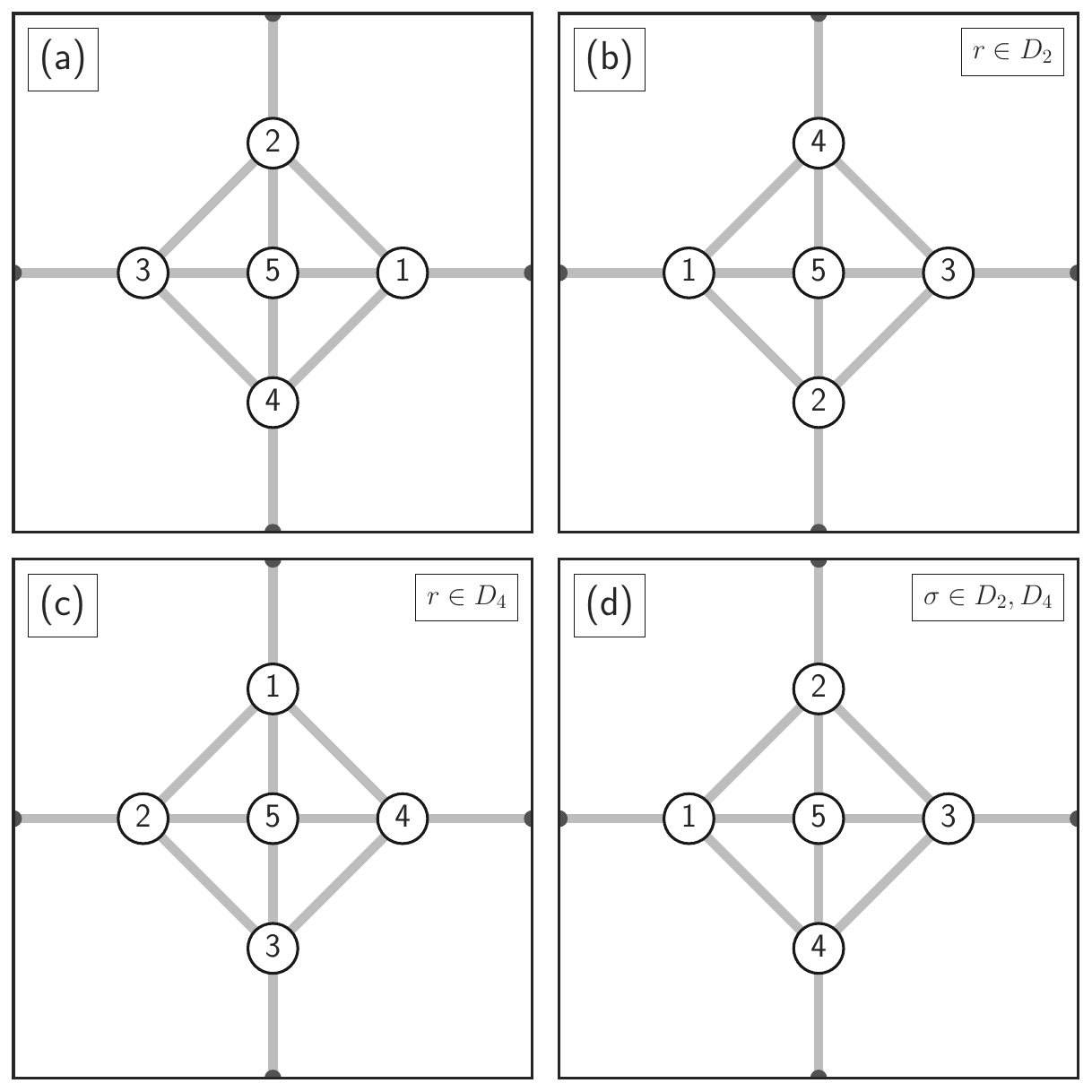}
\caption{
Generators of the dihedral groups $D_2$ and $D_4$ on the diamond network with label assignments for the nodes shown in (a).
}
\label{fig:gensdiamond}
\end{figure}

In Fig.~\ref{fig:gensdiamond}, we depict the action of these generators on the nodes of the network. In Fig.~\ref{fig:gensdiamond}(a), we assign labels to the nodes of this network.
The generators $r$ and $\sigma$ then permute these nodes as shown in Figs.~\ref{fig:gensdiamond}(c) and (d), respectively.

In Fig.~\ref{fig:gensdiamond}(a) we assign labels to the nodes of this network.
As shown in Fig.~\ref{fig:gensdiamond}(c), the generator $r$ corresponds to a $90^\circ$ counterclockwise rotation of the network about the center node.
The generator $\sigma$, shown in Fig.~\ref{fig:gensdiamond}(d), corresponds to a horizontal reflection about the center.

The character table for $D_4$ is shown in Table.~\ref{tab:D4irreps}. 
Each element of the character table contains the character $\chi^\Gamma = \tr \bD^\Gamma(g)$ for a group element $g$ in the conjugacy class $\mathcal{C}$.

The matrix representations we use for the one-dimensional IRREPs $\Gamma =1$, $1'$, $1''$, and $1'''$ are the same as the characters values in reported in Table.~\ref{tab:D4irreps}, i.e., $\bD^\Gamma(g) = \chi^\Gamma(g)$. For the two-dimensional IRREP $\Gamma = 2$, we use the following matrices for the generators:
\begin{equation}
\bD^2(r) = \mqty(0 & -1\\1 & 0) \qqc \bD^2(\sigma) = \mqty(-1 & 0 \\ 0 & 1).
\end{equation}

To construct node space representation $\bDu(g)$ we use to analyze the mode structure of the Hessian under $D_4$, we first define a $5$-dimensional matrix representation $\bD^{(p)}(g)$ using permutation matrices of the nodes as shown in Fig.~\ref{fig:gensdiamond}.  
We then construct the node space representation by taking the tensor product of $\bD^{(p)}(g)$ and the representation matrices of the $\Gamma=2$ IRREP,
\begin{equation}
\bDu(g) = \bD^{(p)}(g) \otimes \bD^{2}(g)\qc \forall g\in G.
\end{equation}

\subsubsection{\texorpdfstring{$D_2$}{D2} Group}\label{sec:D2}

The symmetry of the elastic response in Figs.~\ref{fig:symbreak}(b) and \ref{fig:levels}(b)  is described by the two-fold dihedral group $D_2$ with group presentation
\begin{equation}
D_2 = \langle r, \sigma \mid r^2=\sigma^2=(\sigma r)^2=1\rangle.\label{eq:presD2}
\end{equation}
The generator $r$ is a $180^\circ$ counterclockwise rotation about the center, while $\sigma$ is a reflection that is the same as $D_4$. 
The action of these generators on the nodes is shown in Fig.~\ref{fig:gensdiamond}(b) and (d).

The character table of $D_2$ is shown in Table~\ref{tab:D2irreps}.
Because all of the IRREPs are one-dimensional, the representation matrices are all the same as the character values, $\bD^\Gamma(g) = \chi^\Gamma(g)$.

To construct the node space representation, we the use fact that $D_2 \subset D_4$ and simply reuse the representation matrices of $D_4$ for the appropriate subset of group elements.

\begin{table}[h]
\centering
\begin{tabular}{c c | c c c c c c}
\hline\hline
Conjugacy Class & Class Size & \multicolumn{5}{c}{IRREP $\Gamma$}\\
$\mathcal{C}$ & $n_{\mathcal{C}}$ & $1$ & $1'$ & $1''$ & $1'''$ & $2$\\\hline
$\{I \}$ & $1$  & $1$  & $1$  & $1$  & $1$ & $2$\\ 
$\{r, r^3 \}$& $2$  & $1$  & $1$  & $-1$  & $-1$ & $0$\\ 
$\{\sigma, \sigma r^2 \}$& $2$  & $1$  & $-1$  & $1$  & $-1$ & $0$\\ 
$\{r^2 \}$ & $1$  & $1$  & $1$  & $1$  & $1$ & $-2$\\
$\{\sigma r, \sigma r^3 \}$& $2$  & $1$  & $-1$  & $-1$  & $1$ & $0$\\ 
\hline\hline
\end{tabular}
\caption{
Character table for $D_4$. 
}\label{tab:D4irreps}
\end{table}

\begin{table}[h]
\centering
\begin{tabular}{c c | c c c c c}
\hline\hline
Conjugacy Class & Class Size & \multicolumn{4}{c}{IRREP $\Gamma$}\\
$\mathcal{C}$ & $n_{\mathcal{C}}$ & $1$ & $1'$ & $1''$ & $1'''$ \\\hline
$\{I \}$ & $1$  & $1$  & $1$  & $1$  & $1$ \\ 
$\{r\}$& $1$  & $1$  & $1$  & $-1$  & $-1$ \\ 
$\{\sigma \}$& $1$  & $1$  & $-1$  & $1$  & $-1$ \\ 
$\{\sigma r \}$& $1$  & $1$  & $-1$  & $-1$  & $1$ \\ 
\hline\hline
\end{tabular}
\caption{
Character table for $D_2$. 
}\label{tab:D2irreps}
\end{table}

\subsection{Periodic Square Lattice}\label{app:lattice}

Here we describe the groups used to capture the symmetries of the diamond network in Figs.~\ref{fig:subgroup} and \ref{fig:levels4x4}.

\subsubsection{\texorpdfstring{$(Z_4\otimes Z_4)\rtimes D_4$}{Z4xZ4xD4} Group}

The geometric symmetry of the $4\times 4$ periodic square lattice is captured by the group $(Z_4\otimes Z_4)\rtimes D_4$ with group presentation
\begin{multline}
(Z_4\otimes Z_4)\rtimes D_4 = \langle r, \sigma, x, y \mid r^4 = \sigma^2 = x^4 = y^4\\ = (\sigma r)^2 = (\sigma x)^2 =1, rx=yr\rangle.
\end{multline}
This group has four generators: $r$ corresponds to a $90^\circ$ counterclockwise rotation, $\sigma$ corresponds to reflection across the central vertical axis, and $x$ and $y$ represent translations by $1/4$ a unit cell along the $x$- and $y$-axes, respectively. 

\begin{figure}[h]
\centering
\includegraphics[width=\linewidth]{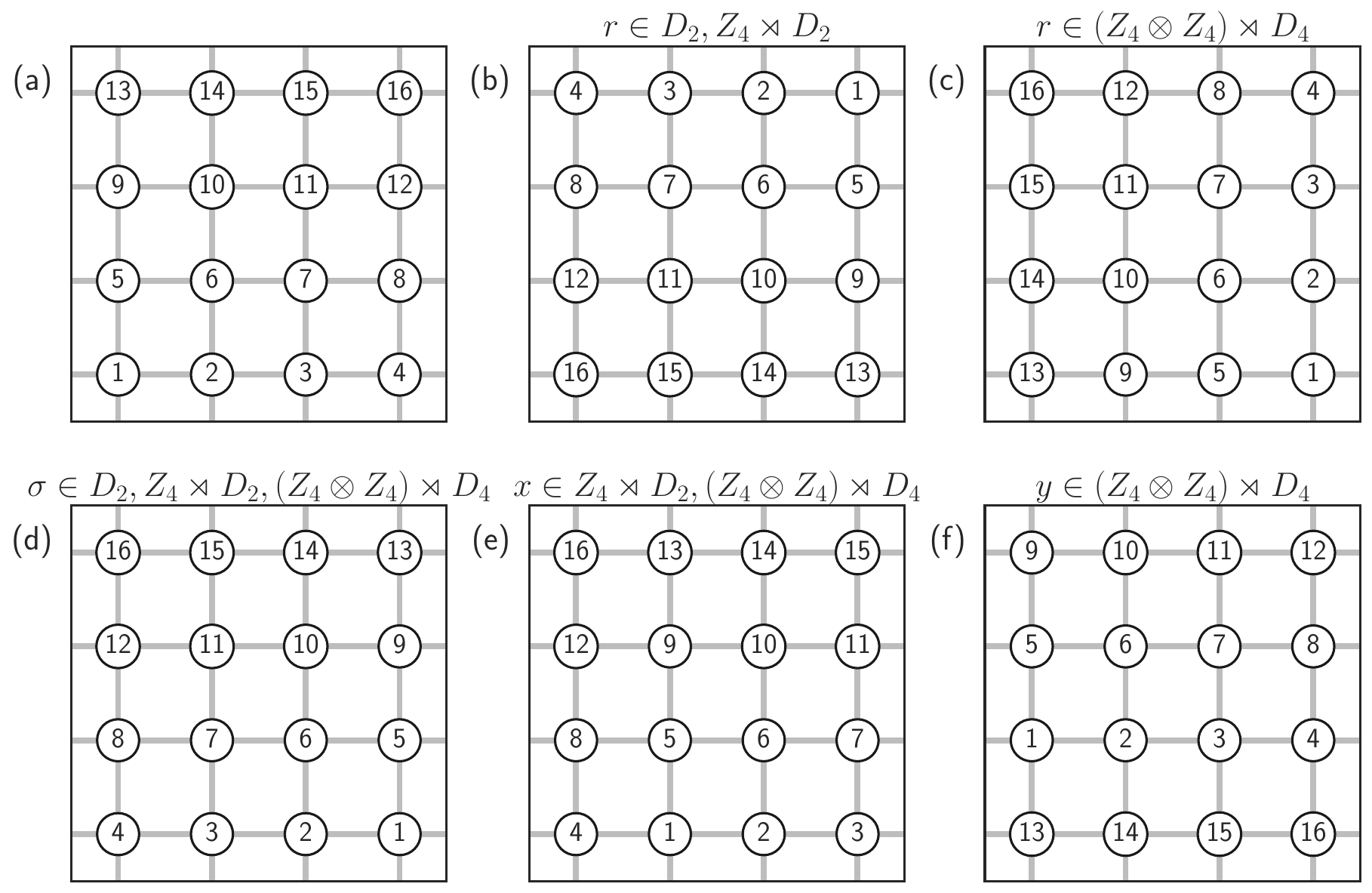}
\caption{
Generators of the groups $(Z_4\otimes Z_4) \rtimes D_4$,  $Z_4 \rtimes D_2$, and $D_2$ the $4\times 4$ square lattice network with label assignments for the nodes shown in (a).
}
\label{fig:gens4x4lattice}
\end{figure}

In Fig.~\ref{fig:gens4x4lattice}, we depict the action of these generators on the nodes of the network, with labels assignments for the nodes shown in Fig.~\ref{fig:gens4x4lattice}(a). We depict the actions of $r$, $\sigma$, $x$, and $y$ in Figs.~\ref{fig:gens4x4lattice}(c)-(f), respectively.

The character table for this group is reported in Table~\ref{tab:Z4Z4D4irreps}. 
This group has 20 conjugacy classes and 20 IRREPs.
We take the matrix representations for the six one-dimensional IRREPs, $1^{(1)}$ through $1^{(6)}$, to simply be the characters for each element. 
For the generators of the six two-dimensional IRREPs, we use the matrix representations
\begin{widetext}
\begin{equation}
\begin{aligned}
\bD^{2^{(1)}}(r) &= \mqty(0 & 1\\ 1 & 0), & \bD^{2^{(1)}}(\sigma) &= \mqty(1 & 0\\ 0 & 1), & \bD^{2^{(1)}}(x) &= \mqty(-1 & 0\\ 0 & 1), & \bD^{2^{(1)}}(y) &= \mqty(1 & 0\\ 0 & -1),\\
\bD^{2^{(2)}}(r) &= \mqty(0 & -1\\ 1 & 0), & \bD^{2^{(2)}}(\sigma) &= \mqty(-1 & 0\\ 0 & 1), & \bD^{2^{(2)}}(x) &= \mqty(1 & 0\\ 0 & 1), & \bD^{2^{(2)}}(y) &= \mqty(1 & 0\\ 0 & 1),\\
\bD^{2^{(3)}}(r) &= \mqty(0 & -1\\ 1 & 0), & \bD^{2^{(3)}}(\sigma) &= \mqty(-1 & 0\\ 0 & 1), & \bD^{2^{(3)}}(x) &= \mqty(-1 & 0\\ 0 & 1), & \bD^{2^{(3)}}(y) &= \mqty(1 & 0\\ 0 & -1),\\
\bD^{2^{(4)}}(r) &= \mqty(0 & -1\\ 1 & 0), & \bD^{2^{(4)}}(\sigma) &= \mqty(-1 & 0\\ 0 & 1), & \bD^{2^{(4)}}(x) &= \mqty(1 & 0\\ 0 & -1), & \bD^{2^{(4)}}(y) &= \mqty(-1 & 0\\ 0 & 1),\\
\bD^{2^{(5)}}(r) &= \mqty(0 & 1\\ -1 & 0), & \bD^{2^{(5)}}(\sigma) &= \mqty(1 & 0\\ 0 & -1), & \bD^{2^{(5)}}(x) &= \mqty(-1 & 0\\ 0 & -1), & \bD^{2^{(5)}}(y) &= \mqty(-1 & 0\\ 0 & -1),\\
\bD^{2^{(6)}}(r) &= \mqty(0 & 1\\ 1 & 0), & \bD^{2^{(6)}}(\sigma) &= \mqty(-1 & 0\\ 0 & -1), & \bD^{2^{(6)}}(x) &= \mqty(1 & 0\\ 0 & -1), & \bD^{2^{(6)}}(y) &= \mqty(-1 & 0\\ 0 & 1),
\end{aligned}
\end{equation}
and for the six four-dimensional IRREPs, we use

\begin{equation}
\begin{aligned}
\bD^{4^{(1)}}(r) &= \mqty(0 & 0 & -1 & 0\\ 0 & 0 & 0 & 1\\ 1 & 0 & 0 & 0\\ 0 & 1 & 0 & 0),& \bD^{4^{(1)}}(\sigma) &= \mqty(-1 & 0 & 0 & 0\\ 0 & 1 & 0 & 0\\ 0 & 0 & 1 & 0\\ 0 & 0 & 0 & 1),&
\bD^{4^{(1)}}(x) &= \mqty(0 & -1 & 0 & 0\\ 1 & 0 & 0 & 0\\ 0 & 0 & 1 & 0\\ 0 & 0 & 0 & 1),& \bD^{4^{(1)}}(y) &= \mqty(1 & 0 & 0 & 0\\ 0 & 1 & 0 & 0\\ 0 & 0 & 0 & -1\\ 0 & 0 & 1 & 0),\\
\bD^{4^{(2)}}(r) &= \mqty(0 & 0 & -1 & 0\\ 0 & 0 & 0 & 1\\ 1 & 0 & 0 & 0\\ 0 & 1 & 0 & 0),& \bD^{4^{(2)}}(\sigma) &= \mqty(-1 & 0 & 0 & 0\\ 0 & 1 & 0 & 0\\ 0 & 0 & 1 & 0\\ 0 & 0 & 0 & 1),&
\bD^{4^{(2)}}(x) &= \mqty(0 & -1 & 0 & 0\\ 1 & 0 & 0 & 0\\ 0 & 0 & -1 & 0\\ 0 & 0 & 0 & -1),& \bD^{4^{(2)}}(y) &= \mqty(-1 & 0 & 0 & 0\\ 0 & -1 & 0 & 0\\ 0 & 0 & 0 & -1\\ 0 & 0 & 1 & 0),\\
\bD^{4^{(3)}}(r) &= \mqty(0 & 0 & 1 & 0\\ 0 & 0 & 0 & -1\\ 1 & 0 & 0 & 0\\ 0 & 1 & 0 & 0),& \bD^{4^{(3)}}(\sigma) &= \mqty(0 & 0 & 1 & 0\\ 0 & 0 & 0 & -1\\ 1 & 0 & 0 & 0\\ 0 & -1 & 0 & 0),&
\bD^{4^{(3)}}(x) &= \mqty(0 & -1 & 0 & 0\\ 1 & 0 & 0 & 0\\ 0 & 0 & 0 & -1\\ 0 & 0 & 1 & 0),& \bD^{4^{(3)}}(y) &= \mqty(0 & 1 & 0 & 0\\ -1 & 0 & 0 & 0\\ 0 & 0 & 0 & -1\\ 0 & 0 & 1 & 0),\\
\bD^{4^{(4)}}(r) &= \mqty(0 & 0 & -1 & 0\\ 0 & 0 & 0 & 1\\ 1 & 0 & 0 & 0\\ 0 & 1 & 0 & 0), & \bD^{4^{(4)}}(\sigma) &= \mqty(0 & 0 & 1 & 0\\ 0 & 0 & 0 & -1\\ 1 & 0 & 0 & 0\\ 0 & -1 & 0 & 0),& 
\bD^{4^{(4)}}(x) &= \mqty(0 & -1 & 0 & 0\\ 1 & 0 & 0 & 0\\ 0 & 0 & 0 & -1\\ 0 & 0 & 1 & 0), & \bD^{4^{(4)}}(y) &= \mqty(0 & 1 & 0 & 0\\ -1 & 0 & 0 & 0\\ 0 & 0 & 0 & -1\\ 0 & 0 & 1 & 0),\\
\bD^{4^{(5)}}(r) &= \mqty(0 & 0 & 1 & 0\\ 0 & 0 & 0 & 1\\ -1 & 0 & 0 & 0\\ 0 & 1 & 0 & 0), & \bD^{4^{(5)}}(\sigma) &= \mqty(-1 & 0 & 0 & 0\\ 0 & -1 & 0 & 0\\ 0 & 0 & 1 & 0\\ 0 & 0 & 0 & -1),& 
\bD^{4^{(5)}}(x) &= \mqty(1 & 0 & 0 & 0\\ 0 & 1 & 0 & 0\\ 0 & 0 & 0 & -1\\ 0 & 0 & 1 & 0), & \bD^{4^{(5)}}(y) &= \mqty(0 & -1 & 0 & 0\\ 1 & 0 & 0 & 0\\ 0 & 0 & 1 & 0\\ 0 & 0 & 0 & 1),\\
\bD^{4^{(6)}}(r) &= \mqty(0 & 0 & -1 & 0\\ 0 & 0 & 0 & 1\\ 1 & 0 & 0 & 0\\ 0 & 1 & 0 & 0), & \bD^{4^{(6)}}(\sigma) &= \mqty(1 & 0 & 0 & 0\\ 0 & -1 & 0 & 0\\ 0 & 0 & -1 & 0\\ 0 & 0 & 0 & -1),& 
\bD^{4^{(6)}}(x) &= \mqty(0 & -1 & 0 & 0\\ 1 & 0 & 0 & 0\\ 0 & 0 & -1 & 0\\ 0 & 0 & 0 & -1), & \bD^{4^{(6)}}(y) &= \mqty(-1 & 0 & 0 & 0\\ 0 & -1 & 0 & 0\\ 0 & 0 & 0 & -1\\ 0 & 0 & 1 & 0).
\end{aligned}
\end{equation}
\end{widetext}

Similar to the construction node space representation for the diamond network in the previous section, we first create a node permutation representation $\bD^{(p)}(g)$, this time of dimension 16, using the permutation matrices of the nodes as shown in Fig.~\ref{fig:gens4x4lattice}.
The node space representation can then be created by taking the outer product of this representation with the two-dimensional IRREP $\Gamma=2^{(2)}$,
\begin{equation}
\bDu(g) = \bD^{(n)}(g) \otimes \bD^{2^{(2)}}(g)\qc \forall g \in G.
\end{equation}

\subsubsection{\texorpdfstring{$Z_4\rtimes D_2$}{Z4xD2} Group}

The symmetry of the elastic response in Figs.~\ref{fig:subgroup}(b) and \ref{fig:levels4x4}(b) is captured by the group $Z_4\rtimes D_2$ with group presentation
\begin{equation}
Z_4\rtimes D_2 = \langle r, \sigma, x, y \mid r^2= \sigma^2 = x^4 = (\sigma r)^2 = (\sigma x)^2 = 1\rangle
\end{equation}
This group has three generators: $r$ corresponds to a $180^\circ$ counterclockwise rotation, $\sigma$ corresponds to reflection across the central vertical axis, and $x$ represents translations by $1/4$ a unit cell along the $x$-axis. 
The action of these generators on the nodes is shown in Figs.~\ref{fig:gens4x4lattice}(b), (d), and (e), respectively.

The character table of $Z_4\rtimes D_2$ is shown in Table~\ref{tab:D2irreps}. 
This group has 10 conjugacy classes and 10 IRREPs.
We take the matrix representations for the eight one-dimensional IRREPs, $1^{(1)}$ through $1^{(8)}$, to simply be the characters for each element. 
For the generators of the two two-dimensional representations, we use the matrix representations
\begin{widetext}
\begin{equation}
\begin{aligned}
\bD^{2}(r) &= \mqty(-1 & 0\\ 0 & 1), & \bD^{2}(\sigma) &= \mqty(-1 & 0\\ 0 & 1), & \bD^{2}(x) &= \mqty(0 & 1\\ -1 & 0),\\
\bD^{2'}(r) &= \mqty(1 & 0\\ 0 & -1), & \bD^{2'}(\sigma) &= \mqty(-1 & 0\\ 0 & 1), & \bD^{2'}(x) &= \mqty(0 & 1\\ -1 & 0).
\end{aligned}
\end{equation}
\end{widetext}

To construct the node space representation, use the fact that ${Z_4\rtimes D_2 \subset (Z_4\otimes Z_4)\rtimes D_4}$ and simply reuse the representation matrices of $(Z_4\otimes Z_4)\rtimes D_4$ for the appropriate subset of group elements.

\subsubsection{\texorpdfstring{$D_2$}{D2} Group}
The symmetry of the elastic response in Figs.~\ref{fig:subgroup}(c) and \ref{fig:levels4x4}(c) is captured by the two-fold dihedral group $D_2$ with group presentation shown in Eq.~\eqref{eq:presD2}.
The action of the two generators $r$ and $\sigma$ are shown in Figs.~\ref{fig:gens4x4lattice}(b) and (d), respectively.
The character table and IRREP matrices are presented in Sec.~\ref{sec:D2}.

To construct the node space representation, use the fact that $D_2 \subset Z_4\rtimes D_2$ and simply reuse the representation matrices of $Z_4\rtimes D_2$ for the appropriate subset of group elements.

\begin{table*}[th]
\centering
\begin{tabular}{c c | c c c c c c c c c c c c c c c c c c c c}
\hline\hline
Conjugacy Class & Class Size &\multicolumn{20}{c}{IRREP $\Gamma$}\\
$\mathcal{C}$ & $n_{\mathcal{C}}$ & $1^{(1)}$ & $1^{(2)}$ & $1^{(3)}$ & $1^{(4)}$ & $1^{(5)}$ & $1^{(6)}$ & $1^{(7)}$ & $1^{(8)}$ & $2^{(1)}$ & $2^{(2)}$ & $2^{(3)}$ & $2^{(4)}$ & $2^{(5)}$ & $2^{(6)}$ & $4^{(1)}$ & $4^{(2)}$ & $4^{(3)}$ & $4^{(4)}$ & $4^{(5)}$ & $4^{(6)}$\\\hline
$\{I\}$ & $1$ & $1$ & $1$ & $1$ & $1$ & $1$ & $1$ & $1$ & $1$ & $2$ & $2$ & $2$ & $2$ & $2$ & $2$ & $4$ & $4$ & $4$ & $4$ & $4$ & $4$\\
$\{r, \cdots\}$ & $16$ & $1$ & $1$ & $1$ & $1$ & $-1$ & $-1$ & $-1$ & $-1$ & $0$ & $0$ & $0$ & $0$ & $0$ & $0$ & $0$ & $0$ & $0$ & $0$ & $0$ & $0$\\
$\{\sigma, \cdots\}$ & $4$ & $1$ & $1$ & $-1$ & $-1$ & $1$ & $1$ & $-1$ & $-1$ & $2$ & $0$ & $0$ & $0$ & $0$ & $-2$ & $2$ & $2$ & $0$ & $0$ & $-2$ & $-2$\\
$\{x, y, \cdots\}$ & $4$ & $1$ & $-1$ & $1$ & $-1$ & $1$ & $-1$ & $1$ & $-1$ & $0$ & $2$ & $0$ & $0$ & $-2$ & $0$ & $2$ & $-2$ & $0$ & $0$ & $2$ & $-2$\\
$\{r^2, \cdots\}$ & $4$ & $1$ & $1$ & $1$ & $1$ & $1$ & $1$ & $1$ & $1$ & $2$ & $-2$ & $-2$ & $-2$ & $-2$ & $2$ & $0$ & $0$ & $0$ & $0$ & $0$ & $0$\\
$\{\sigma r, \cdots\}$ & $8$ & $1$ & $1$ & $-1$ & $-1$ & $-1$ & $-1$ & $1$ & $1$ & $0$ & $0$ & $0$ & $0$ & $0$ & $0$ & $0$ & $0$ & $2$ & $-2$ & $0$ & $0$\\
$\{r x, r y, \cdots\}$ & $16$ & $1$ & $-1$ & $1$ & $-1$ & $-1$ & $1$ & $-1$ & $1$ & $0$ & $0$ & $0$ & $0$ & $0$ & $0$ & $0$ & $0$ & $0$ & $0$ & $0$ & $0$\\
$\{\sigma x, \cdots\}$ & $4$ & $1$ & $-1$ & $-1$ & $1$ & $1$ & $-1$ & $-1$ & $1$ & $0$ & $0$ & $2$ & $-2$ & $0$ & $0$ & $2$ & $-2$ & $0$ & $0$ & $-2$ & $2$\\
$\{x^2, y^2\}$ & $2$ & $1$ & $1$ & $1$ & $1$ & $1$ & $1$ & $1$ & $1$ & $2$ & $2$ & $2$ & $2$ & $2$ & $2$ & $0$ & $0$ & $-4$ & $-4$ & $0$ & $0$\\
$\{\sigma y, \cdots\}$ & $8$ & $1$ & $-1$ & $-1$ & $1$ & $1$ & $-1$ & $-1$ & $1$ & $0$ & $0$ & $-2$ & $2$ & $0$ & $0$ & $0$ & $0$ & $0$ & $0$ & $0$ & $0$\\
$\{x y, \cdots\}$ & $4$ & $1$ & $1$ & $1$ & $1$ & $1$ & $1$ & $1$ & $1$ & $-2$ & $2$ & $-2$ & $-2$ & $2$ & $-2$ & $0$ & $0$ & $0$ & $0$ & $0$ & $0$\\
$\{r^2 x, r^2 y,  \cdots\}$ & $8$ & $1$ & $-1$ & $1$ & $-1$ & $1$ & $-1$ & $1$ & $-1$ & $0$ & $-2$ & $0$ & $0$ & $2$ & $0$ & $0$ & $0$ & $0$ & $0$ & $0$ & $0$\\
$\{\sigma r x, \sigma r y, \cdots\}$ & $16$ & $1$ & $-1$ & $-1$ & $1$ & $-1$ & $1$ & $1$ & $-1$ & $0$ & $0$ & $0$ & $0$ & $0$ & $0$ & $0$ & $0$ & $0$ & $0$ & $0$ & $0$\\
$\{\sigma x y, \cdots\}$ & $8$ & $1$ & $1$ & $-1$ & $-1$ & $1$ & $1$ & $-1$ & $-1$ & $-2$ & $0$ & $0$ & $0$ & $0$ & $2$ & $0$ & $0$ & $0$ & $0$ & $0$ & $0$\\
$\{x^2 y, x y^2, \cdots\}$ & $4$ & $1$ & $-1$ & $1$ & $-1$ & $1$ & $-1$ & $1$ & $-1$ & $0$ & $2$ & $0$ & $0$ & $-2$ & $0$ & $-2$ & $2$ & $0$ & $0$ & $-2$ & $2$\\
$\{\sigma y^2, \cdots\}$ & $4$ & $1$ & $1$ & $-1$ & $-1$ & $1$ & $1$ & $-1$ & $-1$ & $2$ & $0$ & $0$ & $0$ & $0$ & $-2$ & $-2$ & $-2$ & $0$ & $0$ & $2$ & $2$\\
$\{\sigma r x^2, \sigma r x y, \sigma r y^2, \cdots\}$ & $8$ & $1$ & $1$ & $-1$ & $-1$ & $-1$ & $-1$ & $1$ & $1$ & $0$ & $0$ & $0$ & $0$ & $0$ & $0$ & $0$ & $0$ & $-2$ & $2$ & $0$ & $0$\\
$\{r^2 x y, \cdots\}$ & $4$ & $1$ & $1$ & $1$ & $1$ & $1$ & $1$ & $1$ & $1$ & $-2$ & $-2$ & $2$ & $2$ & $-2$ & $-2$ & $0$ & $0$ & $0$ & $0$ & $0$ & $0$\\
$\{\sigma x y^2, \cdots\}$ & $4$ & $1$ & $-1$ & $-1$ & $1$ & $1$ & $-1$ & $-1$ & $1$ & $0$ & $0$ & $2$ & $-2$ & $0$ & $0$ & $-2$ & $2$ & $0$ & $0$ & $2$ & $-2$\\
$\{x^2 y^2\}$ & $1$ & $1$ & $1$ & $1$ & $1$ & $1$ & $1$ & $1$ & $1$ & $2$ & $2$ & $2$ & $2$ & $2$ & $2$ & $-4$ & $-4$ & $4$ & $4$ & $-4$ & $-4$\\
\hline\hline
\end{tabular}
\caption{
Character table for $(Z_4\otimes Z_4)\rtimes D_4$.
}\label{tab:Z4Z4D4irreps}
\end{table*}

\begin{table*}[th]
\centering
\begin{tabular}{c c | c c c c c c c c c c}
\hline\hline
Conjugacy Class & Class Size & \multicolumn{10}{c}{IRREP $\Gamma$}\\
$\mathcal{C}$ & $n_{\mathcal{C}}$ & $1^{(1)}$ & $1^{(2)}$ & $1^{(3)}$ & $1^{(4)}$ & $1^{(5)}$ & $1^{(6)}$ & $1^{(7)}$ & $1^{(8)}$ & $2$ & $2'$\\\hline
$\{I\}$ & $1$ & $1$ & $1$ & $1$ & $1$ & $1$ & $1$ & $1$ & $1$ & $2$ & $2$\\
$\{r, r x^2\}$ & $2$ & $1$ & $1$ & $-1$ & $-1$ & $1$ & $1$ & $-1$ & $-1$ & $0$ & $0$\\
$\{\sigma, \sigma x^2\}$ & $2$ & $1$ & $1$ & $1$ & $1$ & $-1$ & $-1$ & $-1$ & $-1$ & $0$ & $0$\\
$\{x, x^3\}$ & $2$ & $1$ & $-1$ & $1$ & $-1$ & $1$ & $-1$ & $1$ & $-1$ & $0$ & $0$\\
$\{\sigma r\}$ & $1$ & $1$ & $1$ & $-1$ & $-1$ & $-1$ & $-1$ & $1$ & $1$ & $2$ & $-2$\\
$\{r x, r x^3\}$ & $2$ & $1$ & $-1$ & $-1$ & $1$ & $1$ & $-1$ & $-1$ & $1$ & $0$ & $0$\\
$\{\sigma x, \sigma x^3\}$ & $2$ & $1$ & $-1$ & $1$ & $-1$ & $-1$ & $1$ & $-1$ & $1$ & $0$ & $0$\\
$\{x^2\}$ & $1$ & $1$ & $1$ & $1$ & $1$ & $1$ & $1$ & $1$ & $1$ & $-2$ & $-2$\\
$\{\sigma r x, \sigma r x^3\}$ & $2$ & $1$ & $-1$ & $-1$ & $1$ & $-1$ & $1$ & $1$ & $-1$ & $0$ & $0$\\
$\{\sigma r x^2\}$ & $1$ & $1$ & $1$ & $-1$ & $-1$ & $-1$ & $-1$ & $1$ & $1$ & $-2$ & $2$\\\hline\hline
\end{tabular}
\caption{
Character table for $Z_4\rtimes D_2$.
}
\end{table*}

\end{document}